\documentclass{IEEEoj}

\usepackage[T1]{fontenc}
\usepackage{fix-cm}
\usepackage{cite}
\usepackage{amsmath,amssymb,amsfonts}
\usepackage{algorithmic}
\usepackage{graphicx,color}
\usepackage{textcomp}
\usepackage{algorithm}
\usepackage{algorithmic}
\usepackage{array}
\usepackage{tabularx}
\usepackage[acronym,,nohypertypes={acronym,notation}]{glossaries}
\newacronym{3d}{3-D}{three-dimensional}
\newacronym{3gpp}{3GPP}{3rd Generation Partnership Project}
\newacronym{5g}{5G}{fifth generation}
\newacronym{6g}{6G}{sixth generation}
\newacronym{ao}{AO}{alternating optimization}
\newacronym{aoa}{AoA}{angle of arrival}
\newacronym{aod}{AoD}{angle of departure}
\newacronym{as}{AS}{antenna selection}
\newacronym{awgn}{AWGN}{additive white gaussian noise}
\newacronym{b5g}{B5G}{beyond fifth generation}
\newacronym[plural = BSs, firstplural = base statios (BSs)]{bs}{BS}{base station}
\newacronym{ber}{BER}{bit-error-rate}
\newacronym{csi}{CSI}{channel state information}
\newacronym{chest}{CHEST}{channel estimation}
\newacronym{ccdf}{CCDF}{complementary cumulative distribution function}
\newacronym{ccm}{CCM}{complex circle manifold}
\newacronym{cs}{CS}{channel sounding}
\newacronym{ccp}{CCP}{convex-concave procedure}
\newacronym{dl}{DL}{downlink}
\newacronym{ee}{EE}{energy efficiency}
\newacronym{em}{EM}{electromagnetic}
\newacronym{er}{ER}{ergodic rate}

\newacronym{fp}{FP}{fractional programming}
\newacronym{ga}{GA}{genetic algorithm}
\newacronym{gf}{GF}{grant-free}
\newacronym{hbf}{HBF}{hybrid beamforming}
\newacronym[plural = HEMs]{hem}{HEM}{{\it heuristic evolutionary}}
\newacronym{iao}{i-AO}{iterative-alternating optimization}
\newacronym{iid}{i.i.d.}{independent and identically distributed}
\newacronym{iot}{IoT}{internet of things}
\newacronym{isomap}{Isomap}{Isometric Mapping}
\newacronym{kkt}{KKT}{Karush–Kuhn–Tucker}
\newacronym[plural = KPIs, firstplural = key performance indicators (KPIs)]{kpi}{KPI}{key performance indicator}
\newacronym{ldt}{LDT}{Lagrangian Dual Transform}
\newacronym{los}{LoS}{line-of-sight}
\newacronym{lle}{LLE}{Locally Linear Embedding}

\newacronym{mimo}{MIMO}{multiple-input multiple-output}
\newacronym{m-mimo}{mMIMO}{massive MIMO}
\newacronym{mmse}{MMSE}{minimum mean squared error}
\newacronym{mrc}{MRC}{maximum ratio combiner}
\newacronym{mcs}{MCs}{Monte-Carlo Simulation}
\newacronym{mrt}{MRT}{maximum ratio transmission}
\newacronym{mu}{MU}{multi-user}
\newacronym{mo}{MO}{Manifold optimization}
\newacronym{mds}{MDS}{Multidimensional Scaling }
\newacronym{mmtc}{mMTC}{machine massive type communications}
\newacronym{mm}{MM}{majorization minimization}
\newacronym{mtc}{MTC}{machine type communications}
\newacronym{ml}{ML}{machine-learning}
\newacronym{np-complete}{NP-complete}{nondeterministic polynomial-time complete}
\newacronym{noma}{NOMA}{non-orthogonal multiple access}
\newacronym{nlos}{NLoS}{non line-of-sight}
\newacronym[plural=NNs]{nn}{NN}{neural network}
\newacronym{of}{OF}{objective function}
\newacronym{olp}{OLP}{optimal linear precoder}
\newacronym{pa}{PA}{power amplifier}
\newacronym{pc}{PC}{pilot contamination}
\newacronym{pbo}{PBO}{passive beamforming optimization}
\newacronym{pbf}{PBF}{passive beamforming}
\newacronym{pca}{PCA}{principal component analysis}
\newacronym{prf}{PRF}{pilot-reuse factor}
\newacronym{pso}{PSO}{particle swarm optimization}
\newacronym{qos}{QoS}{Quality of Service}
\newacronym{rf}{RF}{radio frequency}
\newacronym{re}{RE}{resource efficiency}
\newacronym{rm}{RM}{Riemannian manifold}
\newacronym[plural=RISs]{ris}{RIS}{reconfigurable intelligent surface}
\newacronym{rcg}{RCG}{Riemann conjugate gradient}
\newacronym{ra}{RA}{random access}
\newacronym{rn}{RN}{Riemannian Newton}
\newacronym{rtr}{RTR}{Riemannian Trust-Region}
\newacronym{rgd}{RGD}{Riemannian Gradient Descent}
\newacronym{se}{SE}{spectral efficiency}
\newacronym{sfp}{SFP}{sequential fractional programming}
\newacronym{sinr}{SINR}{signal-to-interference-plus-noise ratio}
\newacronym{snr}{SNR}{signal-to-noise ratio}
\newacronym{sr}{SR}{sum rate}
\newacronym{sdr}{SDR}{semidefinite relaxation}
\newacronym{sca}{SCA}{sucessive convex approximation}
\newacronym{socp}{SOCP}{second-order cone programming }
\newacronym{sll}{SLL}{side lobe level}
\newacronym{sdp}{SDP}{semidefinite programming}
\newacronym{sg}{SG}{sub-gradient}
\newacronym{sa}{SA}{simulated annealing }
\newacronym{tdd}{TDD}{time-division duplex}
\newacronym{t-sne}{t-SNE}{t-Distributed Stochastic Neighbor Embedding}
\newacronym[plural=UEs, firstplural=users' equipment (UEs)]{ue}{UE}{user's equipment}
\newacronym{ul}{UL}{uplink}
\newacronym{ula}{ULA}{uniform linear array}
\newacronym{upa}{UPA}{uniform planar array}
\newacronym{uspa}{USPA}{uniform squared planar array}
\newacronym{ut}{UT}{unit terminal}
\newacronym{umap}{UMAP}{Uniform Manifold Approximation and Projection}
\newacronym{vr}{VR}{visibility region}
\newacronym{wbd}{WBD}{Wide-beam design}
\newacronym{xl-mimo}{XL-MIMO}{extra-large scale massive MIMO}
\newacronym{zf}{ZF}{Zero-Forcing}
\newacronym{zft}{ZFT}{zero-forcing transmission}

\usepackage[colorlinks=true, urlcolor=blue, linkcolor=blue, citecolor=red]{hyperref}
\usepackage{subcaption}

\definecolor{gray}{rgb}{0.5,0.5,0.5}

\newcommand{\colo}{\textcolor{ojcscol}}

\newcommand{\dm} {\, \colo{$\diamond$}\, }
\renewcommand{\algorithmiccomment}[1]{{\color{gray} // #1}}

\def\BibTeX{{\rm B\kern-.05em{\sc i\kern-.025em b}\kern-.08em
T\kern-.1667em\lower.7ex\hbox{E}\kern-.125emX}}
\AtBeginDocument{\definecolor{ojcolor}{cmyk}{0.93,0.59,0.15,0.02}}
\def\OJlogo{\vspace{-4pt}\hskip-4pt\includegraphics[height=18pt]{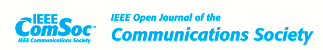}}

\newcommand{\ta}{\textcolor{black}}

\begin{document}
\receiveddate{XX Month, XXXX}
\reviseddate{XX Month, XXXX}
\accepteddate{XX Month, XXXX}
\publisheddate{XX Month, XXXX}
\currentdate{11 January, 2024}
\doiinfo{OJCOMS.2024.011100}

\title{Manifold-Based Optimizations for RIS-Aided Massive MIMO Systems}

\author{ 
\uppercase{Wilson de Souza Junior} \IEEEauthorrefmark{1},
\uppercase{David William Marques Guerra} \IEEEauthorrefmark{1}, 
\uppercase{José Carlos Marinello} \IEEEauthorrefmark{2}, 
\uppercase{Taufik Abrão}, \IEEEauthorrefmark{1},  and
\uppercase{Ekram Hossain} \IEEEauthorrefmark{3}
}

\affil{State University of Londrina  (UEL), CEP: 86057-970, Londrina, PR, Brazil.}

\affil{Federal University of Technology - Parana (UTFPR), CEP: 86300-000, Cornélio Procópio, PR, Brazil.}

\affil{Department of Electrical and Computer Engineering at the University of Manitoba, Winnipeg, Canada.}

\corresp{CORRESPONDING AUTHOR: EKRAM HOSSAIN (e-mail: Ekram.Hossain@umanitoba.ca).}

\authornote{This work was supported in part by the National Council for Scientific and Technological Development (CNPq) of Brazil, Grant: 310681/2019-7, by CAPES, Grant: FC001, in part by Londrina State University (UEL), Brazil, and in part by a Discovery Grant from the Natural Sciences and Engineering Research Council of Canada (NSERC).
}

\markboth{Preparation of Papers for IEEE OPEN JOURNALS}{Author \textit{et al.}}

\begin{abstract}
\gls{mo} is a powerful mathematical framework that can be applied to optimize functions over complex geometric structures, which is particularly useful in advanced wireless communication systems, such as \gls{ris}-aided \gls{m-mimo} and \gls{xl-mimo} systems. \gls{mo} provides a structured approach to tackling complex optimization problems. By leveraging the geometric properties of the manifold, more efficient and effective solutions can be found compared to conventional optimization methods.  This paper provides a tutorial on \gls{mo} technique and provides some applications of \gls{mo} in the context of wireless communications systems. In particular, to corroborate the effectiveness of \gls{mo} methodology, we explore five application examples in  \gls{ris}-aided \gls{m-mimo} system, focusing on fairness, \gls{ee} maximization, intra-cell pilot reuse interference mitigation, and \gls{gf} \gls{ra}. 
\end{abstract}

\begin{IEEEkeywords}
Manifold Optimization (MO), Reconfigurable Intelligent Surfaces (RIS), Massive MIMO (mMIMO), Energy Efficiency, Grant-free Random Access.
\end{IEEEkeywords}

\maketitle


\section{INTRODUCTION}

\IEEEPARstart{M}{ethods} based on \gls{mo} are rooted in their ability to handle complex optimization problems more efficiently and effectively than conventional methods, especially in the context of \gls{m-mimo} systems, especially for \gls{5g} and \gls{b5g} communications such as \gls{6g} systems. 

\begin{table*}[!htbp]
\centering
\caption{Examples of wireless communication problems solved efficiently by using MO techniques}
\label{tab:mo_applications}
\begin{tabularx}{\textwidth}{|>{\raggedright\arraybackslash}X|>{\raggedright\arraybackslash}X|>{\raggedright\arraybackslash}X|}
\hline
\textbf{Wireless example} & \textbf{Problem} & \textbf{MO-based solution} \\ \hline
\gls{ris} Phase Shift Optimization \cite{9802804,10130707,9219206,9154337,9405423,9013288,9681803,10008751,9472958} &   
Optimizing the phase shifts of \gls{ris} elements to maximize some \gls{kpi}. &   
Treating the phase shifts as points on a complex circle manifold allows for more natural and efficient optimization than traditional methods that might struggle with the unit-modulus constraint. \\ \hline

Active beamforming in \gls{m-mimo} Systems \cite{9802804,10130707} &   
Optimize the beams to maximize signal strength and minimize interference. &   
Using \gls{mo} to represent the beam directions on a particular manifold can lead to better beam selection and alignment performance. \\ \hline

Resource Allocation in \gls{b5g} Systems \cite{RA_MO1} &   
Allocating resources such as power and bandwidth to users in a way that maximizes overall system performance. &   
Manifold learning techniques can identify clusters of users or resources, simplifying the optimization problem and leading to more efficient allocation strategies. \\ \hline
\end{tabularx}
\end{table*}

Many optimization problems in wireless communications involve non-convex constraints, such as those arising from the geometry of the problem space. For instance,  in \gls{ris}-aided systems, specifically for entirely passive \gls{ris}, the imposed phase shifts by the \gls{ris}, lie on a unit circle in the complex plane, which forms a non-convex constraint.  \gls{mo} can treat these constraints more naturally by considering the problem as an optimization over a smooth manifold, exploiting {\it geometric properties} of the problem, and allowing a more efficient solution. This is done by leveraging some ideas, such as the {\it local linearization}, which is a technique that benefits from the ability to linearize the problem space locally,  e.g., manifolds are locally Euclidean, meaning they can be linearized around any point, allowing the use of linear optimization techniques such as gradient descent and Newton's method in a more generalized form.  Another motivation for using manifold-based optimizations is the ability to handle {\it high-dimensional data}, often present in wireless communication systems, making optimization challenging. For instance, manifold learning techniques can transform high-dimensional data into a more manageable form, improving optimization in signal processing and resource allocation contexts. Finally, \gls{mo} methods offer versatility to deal with constraints and symmetries, founded in optimization problems that are difficult to handle with traditional optimization methods; for example, \gls{mo} can handle constraints like {\it orthonormality, low rank, positivity}, and even {\it invariance} under group actions, which are common in wireless communication problems. 

\subsection{BACKGROUND}

\glspl{ris} are typically employed to enhance the channel gain between the \gls{bs} and the \gls{ue}, especially when environmental obstacles blockage direct communication. The passive mode of operation, where the device lacks active \gls{rf} components for energy efficiency purposes, is the most desirable scenario for \glspl{ris} since, in this mode, the \gls{ris} does not perform any sophisticated processing on it, reducing its complexity and eliminating eventual overhead in the communication protocol. In this sense, the practical problem of how to configure/optimize the \gls{ris} in wireless communications scenarios emerges, imposing some ideas over its potential and physical constraints.
\vspace{2mm}

\noindent\textbf{\gls{wbd} for \gls{ris}-aided systems}: \gls{wbd} for \gls{ris}-aided design refers to the configuration of \gls{ris} to create broad, uniform beams that cover larger areas compared to narrow, highly focused beams. This design is crucial when dealing with \gls{mu}, especially in \gls{mmtc} where many devices are randomly accessing the system. The advantageous features of the \gls{wbd} for \gls{ris} passive beamforming include:
\begin{itemize}
    \item[\dm]{\it Uniform Coverage}, ensuring that all areas within the specified range receive adequate signal strength, reducing the chances of dead zones.
    
    \item[\dm]{\it Simplified Access}: devices can access the network more easily since they do not need to be precisely aligned with a narrow beam.
    
    \item[\dm]{\it Reduced Complexity} simplifies the control and management of the beams, as fewer adjustments are needed to maintain coverage.
    
    \item[\dm]{\it Support for Mobility}: it better accommodates mobile users as the broad beam can maintain connectivity without frequent reconfiguration.

    \item[\dm]{\it Improved Access Success Rate}: with uniform, wide-beam coverage, more devices are likely to successfully access the network on their first attempt, reducing collisions and the need for retransmissions.
    
    \item[\dm]{\it Load Balancing}: Wide-beam can distribute the access load more evenly across the coverage area, preventing congestion and improving overall network efficiency.
    
    \item[\dm]{\it \gls{ee}}: devices can transmit at lower power levels as they benefit from the enhanced signal propagation provided by \gls{ris}, extending battery life.
    
    \item[\dm]{\it Reduced Latency}: By decreasing the probability of access failures and re-transmissions, wide beam design can help reduce access delays, which is critical for time-sensitive \gls{iot} applications.
\end{itemize}

Given the promising gains offered by \glspl{ris}, their deployment is becoming increasingly essential. However, to achieve the conditions necessary for passive \gls{ris} operation, the reflection coefficients must have a unit modulus. This requirement introduces a non-convex constraint, making the optimization process more challenging. To address this, manifold-based algorithms emerge as a promising solution, offering an appealing trade-off between performance and complexity. These algorithms can navigate the non-convex surfaces effectively, ensuring that the unit modulus constraint is met while optimizing the overall system performance.

On this hand, some works approached the manifold-based algorithms for solving different problems related to the \gls{ris}-assisted communications systems. Table \ref{tab:mo_applications} highlights three wireless optimization examples deploying \gls{mo} approach. In \cite{9802804}, the authors proposed a method to maximize \gls{ee} by jointly optimizing the beamforming at the \gls{ris} and \gls{bs}. The \gls{rcg} method was utilized to find solutions on the sphere for active beamforming and on the \gls{ccm} for passive beamforming. Similarly, in \cite{10130707}, the authors used the \gls{rcg} method to optimize both active and passive precoding. However, this paper employed the \gls{ccm} and oblique manifold. In \cite{10008751}, the authors aimed to optimize passive beamforming to nullify interference between the \glspl{ue} completely. They demonstrated that the solution for this purpose lies on the Stiefel manifold. Additionally, several works adopted the \gls{rcg} method to optimize \gls{ris} passive beamforming aiming to maximize some \gls{kpi} in \gls{ris}-enabled systems \cite{9219206,9154337,9405423,9013288,9681803}. {Reference \cite{RA_MO1} provides insights into how \gls{mo} techniques are applied to solve resource allocation problems in \gls{b5g} systems, i.e.,  near-field resource allocation for \gls{xl-mimo} systems, by comparing different methodologies and optimization tools for the beamforming design, including Riemannian \gls{mo}, \gls{ao}, reinforcement learning, and a generative artificial intelligence-based method.

\vspace{2mm}

\noindent\textbf{Integrating \gls{ris} and \gls{noma} in \gls{m-mimo} systems}: The \gls{mu} \gls{ris}-aided \gls{ul} \gls{noma} design is challenging since it requires joint optimization of both the transmit powers and the \gls{ris} reflection coefficients. Because the optimization variables are linked, the problem cannot be solved with a closed-form approach; also, it is a non-convex problem. 

Existing solutions in the literature for \gls{ul} \gls{ris}-\gls{noma} design use the \gls{sdp} method to turn the optimization problem into a convex form that can be solved with convex programming tools. However, \gls{sdp}-based approaches have high complexity, of order $\mathcal{O}(N^{3.5})$, especially when the number of \gls{ris} elements $N$ is high, which renders these solutions unsuitable for \gls{iot} applications, where the \gls{bs} may have limited power resources that cannot accommodate very complex tasks. 

To reduce the complexity, the authors of \cite{9472958} suggest a simple but effective alternating \gls{mo} algorithm that works well for \gls{iot} applications that run on batteries. 
An alternating \gls{mo}-based algorithm has been proposed for solving the phase shifts, analog beamforming, and transmit beamforming in \gls{ris}-aided \gls{ul} \gls{noma}, and compared with the \gls{sca}-based method. Numerical results reveal that the proposed manifold alternating \gls{mo}-based algorithm outperforms the existing schemes when the sum rate is optimized. A \gls{mo} approach aims to provide a low-complexity design with powerful performance. 

In \cite{Li2022_NomaMO}, the authors investigate the role of \gls{ris} in enhancing the sum throughput for \gls{noma} \gls{iot} networks. It proposes a novel resource allocation strategy that optimizes both time allocation and phase shift matrices during wireless energy transfer and wireless information transfer phases, respectively. The strategy employs {\it elements collaborative approximate} and {\it manifold space gradient descent} algorithms for optimization, demonstrating significant performance gains in simulated environments when compared to networks without \gls{ris} or using other resource allocation methods.

Authors in \cite{Papadias2024} explore \gls{ris}-aided \gls{noma} in the \gls{ul} of energy-limited networks. Two optimization problems are addressed: minimizing users' transmit power and maximizing \gls{ee}. Joint optimization of users' transmit powers and \gls{ris} beamforming coefficients is achieved using a novel low-complexity algorithm on a \gls{ccm}. To solve them, the author deploys iterative \gls{ao} algorithms in two steps to jointly optimize the transmit powers of the users and the phase shifts at \gls{ris}, under transmit power and \gls{qos} constraints: \textit{\textbf{i}})  passive beamforming coefficients are used to solve the transmit power optimization problem. \textit{\textbf{ii}}) fixes users' transmit powers and then solves the \gls{ris} coefficients optimization problem. Compared to three conventional \gls{sdp}-based benchmarks, the proposed \gls{mo}-based algorithm demonstrates better performance with reduced computational complexity, particularly when user target data rates are high.  

\subsection{MOTIVATION AND CONTRIBUTION}
This paper discusses the difficulties of finding the best solutions to non-convex problems in \gls{ris}-assisted \gls{m-mimo} systems, mainly those that arise because passive \gls{ris} elements have a unit modulus constraint. Also, this work provides a more efficient and practical approach to solving complex optimization problems in advanced wireless communication systems compared to conventional methods. Hence, we need to leverage the geometric properties of manifolds for better solutions to wireless communication optimization problems. Finally, we address handling high-dimensional data often present in wireless communication systems.

The paper's contributions are as follows:
\begin{itemize}
\item[\dm]  The paper provides a tutorial on \gls{mo} techniques and their applications in wireless communications systems.
\item[\dm]It explores five application examples in \gls{ris}-aided \gls{m-mimo} systems, focusing on a) fairness; b) \gls{ee} maximization; c) intra-cell pilot reuse interference mitigation;
d) grant-free random access

\item[\dm]The paper demonstrates the effectiveness of \gls{mo} methodology in these specific applications.

\item[\dm]It provides a framework for applying \gls{mo} to wireless communication problems, including steps for identifying appropriate manifolds and formulating problems.

\item[\dm]The paper compares \gls{mo} methods with alternative optimization techniques, highlighting \gls{mo}'s advantages in handling non-convex constraints and exploiting geometric properties.

\item[\dm]It presents a catalog of gradient descent-based algorithms adapted for manifold optimization.

\item[\dm]The paper provides detailed case studies on real-life problems using \gls{mo} in \gls{ris}-aided \gls{m-mimo} situations.
\end{itemize}
The motivations and contributions of this paper are significant in advancing the understanding and application of manifold optimization techniques in the context of modern wireless communication systems, particularly those involving \gls{ris} and \gls{m-mimo} technology.

\color{black}

\subsection{ORGANIZATION OF THE PAPER}

{
The remaining content of the article is organized as follows. Section \ref{sec:II} provides an overview of the manifold concept with alternatives for its application and its fundamental tools, and lists a bunch of usual manifolds. Section \ref{sec:III} structures the steps and selection of suitable manifold learning techniques, identifying key constraints and establishing the geometric structure of these constraints as manifolds. Section \ref{sec:IV} discusses the main steps in formulating and solving wireless communication problems using the  \gls{mo} framework. Section \ref{sec:V} develops a practical optimization example in \gls{ris}-aided \gls{m-mimo} systems; it also provides a catalog of gradient descent-based algorithms. Section \ref{sec:VI} provides a detailed case study on four real-life problems using \gls{mo}. In this central part of the paper, four real-life uses of \gls{mo} in \gls{ris}-aided \gls{m-mimo} situations are discussed in detail. These include making networks fairer, making \gls{iot} systems more energy efficient, and allowing intra-cell pilot reuse. The \gls{mo}-based complete solutions for the four real applications in wireless \gls{ris}-aided \gls{m-mimo} systems are provided.} Finally, Section \ref{sec:VII} draws the main conclusion and perspectives on the \gls{mo} for wireless communication applications.
The paper concludes with a summary of the findings and suggestions for future research directions.

\section{MANIFOLD FUNDAMENTALS} \label{sec:II}

In this section, we start by highlighting different alternatives to the \gls{mo} technique. In the subsequent subsection, we explain the manifold optimization framework. Finally, we catalog a list of manifolds found in many different real-world problems.

\subsection{ALTERNATIVES TO MANIFOLD OPTIMIZATION TECHNIQUE}

There are some alternatives for solving non-convex optimization problems, including \glspl{hem}, {\it convex relaxation} techniques, \gls{ml}-based algorithms, and {\it gradient-based} methods. {\bf HEMs}, such as \gls{ga}, \gls{pso}, and \gls{sa}, among others, are capable of performing a global search and are less likely to get trapped in local minima, therefore, being suitable to be applied to a wide range of problems without requiring gradient information, however, they present demerits of {\it a}) computationally intensive, often requiring many function evaluations, making them computationally expensive; {\it b}) lack of guarantees of convergence to the global optimum and can be slow to converge.

The merits of {\bf convex relaxation} techniques, such as \gls{sdp}, and \gls{ccp}, include {\it a}) rigorous framework for approximating non-convex problems into convex ones, and b) polynomial-time solvability. However, these techniques suffer scalability issues, rapidly becoming computationally infeasible for large-scale problems; moreover, the quality of the solution provided by \gls{sdp} and \gls{ccp} depends on how well the non-convex problem can be approximated by a convex one.

Besides,  {\bf \gls{ml}-based algorithms} can present impressive results since it can deal with large-scale problems, providing sub-optimal solutions. However, their feasibility in real-world scenarios is often limited. This limitation arises because many \gls{ml}-based algorithms require an offline training stage (particularly \glspl{nn} in supervised learning), utilizing data collected from real-world scenarios, however, it can be incompatible with the highly dynamic nature of wireless communication environments. The necessity for constant adaptation in these environments makes it challenging to rely on pre-trained models. Therefore, their practical application in wireless communications remains constrained by these real-world considerations.

Finally, {\bf gradient-based} methods, such as {\it gradient descent, Newton's method}, and {\it conjugate gradient}, represent a competitive alternative to the \gls{mo} approach for problems where gradient information is available, revealing strong local convergence properties in such scenarios. However, gradient-based methods can easily get trapped in local minima. As a substantial limitation, these methods require the \gls{of} to be differentiable, which can not be always practical.

The key {\bf pros} and {\bf cons} of \gls{mo} over traditional optimization methods in wireless communications are summarized in Table \ref{table:comparison}, and include a) natural handling of non-convex constraints; b) geometric property exploitation; c) local linearization; d) versatility in handling constraints and symmetries; and e) high-dimensional data management: Wireless communication systems often deal with high-dimensional data. Therefore, \gls{mo} methods can transform this data into a more manageable form, improving signal processing and resource allocation optimization.

\begin{table*}[ht!]
\caption{Comparison of \gls{mo} methods to its alternatives}\label{table:comparison}
\centering
\begin{tabularx}{\textwidth}{|>{\raggedright\arraybackslash}p{2.6cm}|>{\raggedright\arraybackslash}p{7.5cm}|>{\raggedright\arraybackslash}p{6.3cm}|}

\hline

\textbf{Method} & \textbf{Pros} & \textbf{Cons} \\

\hline\hline

\textbf{\gls{mo}} 

& 

{\it Handling Non-Convex Constraints}: \gls{mo} methods naturally handle non-convex constraints by treating the problem as an optimization over a smooth manifold.&  {\it Complexity in Implementation}: Implementing MO methods can be complex due to the need for specialized knowledge in differential geometry and manifold theory.\\
\cline{2-3}
& 
{\it Exploiting Geometric Properties}: \gls{mo} methods leverage the geometric properties of the problem, handling constraints and symmetries (orthonormality, low rank, positivity, and invariance), allowing for more efficient solutions. 
&  

{\it Algorithmic Design Challenges}: The manifold constraint adds complexity to the algorithmic design and theoretical analysis.\\

\cline{2-3}

& {\it Local Linearization}: Manifolds are locally Euclidean, enabling linear optimization techniques in a more generalized form. & 
{\it Computational Overhead}: While efficient, MO methods can still be computationally intensive, especially for high-dimensional problems.\\
\cline{2-3}

 & {\it High-Dimensional Data Management}: MO methods can transform high-dimensional data into a more manageable form, improving optimization in tasks like signal processing and resource allocation. & \\
 
\hline\hline

\textbf{\glspl{hem}} 

& {\it Global Search Capability}: These methods perform a global search and are less likely to get trapped in local minima. 
&  

{\it Computationally Intensive}: They often require many function evaluations, making them computationally expensive. 
\\

\cline{2-3}

& {\it Flexibility}: They can be applied to various problems without requiring gradient information. 
&
{\it Lack of Guarantees}: No guarantee to converge to the global optimum or can converge slowly. \\

\hline\hline
\textbf{Convex Relaxation Techniques} & 
{\it Mathematical Rigor}: These methods provide a rigorous framework for approximating non-convex problems. & 
{\it Approximation Quality}: The quality of the solution depends on how well the non-convex problem can be approximated by a convex one. 
\\

\cline{2-3}

& {\it Polynomial-Time Solvability}: Convex problems can be solved efficiently using polynomial-time algorithms. & 
{\it Scalability Issues}: These methods can become computationally infeasible for large-scale problems. \\
\hline\hline

\textbf{\gls{ml}-based algorithms}
& \ta{\textit{Adaptability}: \gls{ml} methods can adapt to various scenarios and data patterns without requiring explicit modeling of the underlying physical processes. }
&   \ta{\textit{Training Data Requirement}: \gls{ml} methods require large amounts of high-quality training data, which may not always be available or easy to obtain.} 
\\

\cline{2-3}

& \ta{\textit{Data-Driven}: \gls{ml} methods leverage large datasets to learn and improve performance over time, making them suitable for environments where data is abundant. } & 
 \ta{\textit{Computational Complexity}: Training \gls{ml} models, especially deep learning models, can be computationally intensive and time-consuming.}  \\
\cline{2-3}
& 
\ta{\textit{Automation}: Once trained, \gls{ml} models can automate complex decision-making processes, reducing the need for manual intervention. }  &  \ta{\textit{Generalization}: \gls{ml} models may struggle to generalize well to unseen scenarios or out-of-distribution data, leading to suboptimal performance.}  \\
\cline{2-3}
& \ta{ \textit{Scalability}: \gls{ml} algorithms can handle high-dimensional data and scale well with the increasing complexity of wireless communication systems.} & \ta{\textit{Interpretability}: Particularly deep \gls{nn}, often act as black boxes, making it difficult to interpret/understand their decision-making processes.} \\

\hline\hline

\textbf{Gradient-Based Methods}
& {\it Efficiency}: These methods are efficient for problems where gradient information is available. 
& {\it Requirement of Smoothness}: These methods require the \gls{of} to be differentiable. \\
\cline{2-3}
& {\it Local Convergence}: They have strong local convergence properties.  & {\it Local Minima}: They can easily get trapped in local minima.\\
\hline
\end{tabularx}
\end{table*}

\subsection{MANIFOLD OPTIMIZATION FRAMEWORK}
In an optimization framework, we consider the search space $\mathcal{S}$ as the set containing all possible answers to our problem, and a cost function $f:\mathcal{S} \rightarrow \mathbb{R}$ which associates a cost $f(x)$ to each element $x$ of $\mathcal{S}$. The goal is to find $x \in \mathcal{S}$ such that $f(x)$ is minimized: 
\begin{equation}
\arg \min_{x\in \mathcal{S}} f(x).
\end{equation}

We occasionally wish to denote the subset of $\mathcal{S}$ for which the minimal cost is attained. We should keep in mind that this set might be empty. 

The Euclidean structure of $\mathbb{R}^n$ and the \gls{of} $f$'s smoothness are irrelevant to the optimization problem's definition. They are merely structures that we should use algorithmically to our advantage. Assuming linearity, the \gls{mo} approach requires smoothness as the key structure to exploit. 

\subsubsection{Optimization Over Smooth Surfaces}

Manifolds are a fundamental concept in mathematics, particularly in geometry and topology. Manifolds provide a generalization of shapes and spaces that locally resemble Euclidean space. Indeed, optimization on manifolds is a versatile framework for continuous optimization. It encompasses optimization over vectors and matrices and allows optimizing over curved spaces to handle constraints and symmetries such as orthonormality, low rank, positivity, and invariance under group actions \cite{manopt2014}.

Let us consider the set $\mathcal{M}$ as a smooth manifold, and the function $f$ is smooth on $\mathcal{M}$. Optimization over such surfaces can be understood as constrained because $x$ is not free to travel in $\mathbb{R}^n$ space but is only allowed to stay on the surface. The favored alternative viewpoint, in this case, is to consider this as unconstrained optimization in a universe where the smooth surface is the only thing that exists. As a result, the generalized Euclidean methods from unconstrained optimization can be applied to the larger class of optimization over smooth manifolds. We require a correct knowledge of gradient and Hessian on smooth manifolds to generalize techniques such as gradient descent and Newton's method. In the linear instance, this requires including an inner product or Euclidean structure. In a more general situation, it is advisable to exploit the property that smooth manifolds are locally linearizable around all points. The linearization at $x$ is the tangent space. Giving each tangent space its inner product\footnote{Varying smoothly with $x$ in a way to be determined precisely.} transforms the manifold into a \gls{rm}, upon which we construct what is known as a Riemannian structure \cite{Boumal_book_2023, manopt2014}.

\subsubsection{Operators on Riemannian Manifold}

\gls{rm}s are mathematical objects that generalize the notion of Euclidean space to more complex and curved geometries. These spaces are foundational in various fields, including optimization, differential geometry, and theoretical physics \cite{Absil2007}. A Riemannian manifold is locally similar to Euclidean space but differs in that it is equipped with a Riemannian metric tensor. This tensor defines the distances and angles between points on the manifold by assigning a positive definite inner product to each tangent space. This inner product allows for the measurement and interpretation of geometric properties such as length, angle, and curvature. Some key definitions and concepts in Riemannian geometry include:

\begin{itemize}
    \item[\dm]Riemannian Gradient ($\nabla_{\mathcal{M}} f(\boldsymbol{x})$): This is the generalization of the gradient from Euclidean space  $\nabla f$ to Riemannian manifolds. Specifically, the Riemannian gradient of a function $f$ on a manifold $\mathcal{M}$ is the projection of the Euclidean gradient onto the tangent space of the manifold at a given point.  
    \begin{equation}
        \nabla_{\mathcal{M}} f(\boldsymbol{x}) = P_{\mathcal{T}_x \mathcal{M}} (\nabla f(\boldsymbol{x})),
    \end{equation}
    where $P_{\mathcal{T}_x \mathcal{M}}$ is the projection operator onto the tangent space $\mathcal{T}_x \mathcal{M}$.   It represents the direction of the steepest ascent of a function $f$ on the manifold $\mathcal{M}$.

\item[\dm]Retraction Operation ($\operatorname{Retr}_{\mathcal{M}}(\boldsymbol{x})$): The retraction operator $\operatorname{Retr}_{\mathcal{M}}(\boldsymbol{x})$ of a point on a manifold $\mathcal{M}$ is the projection of the given point $\boldsymbol{x}$ over the manifold $\mathcal{M}$. Retractions are used to ensure that optimization steps remain on the manifold.
\end{itemize}

The Riemannian gradient and retraction operation are essential for algorithms that optimize manifolds, as they ensure that the iterative steps respect the manifold's geometric structure. Moreover, we should bear in mind that each manifold has its own projection operator on the tangent space, as well as the retraction operator.

\subsubsection{Challenges in Manifold Optimization}

If additional constraints other than the manifold constraint are applied, one can add an indicator function of the feasible set of such additional constraints in the \gls{of}. Hence, the optimization problem covers a general formulation for \gls{mo}. Moreover, the manifold constraint is one of the main difficulties in algorithmic design and theoretical analysis.

One of the main challenges in \gls{mo} usually is the non-convexity of the manifold constraints. By utilizing the geometry of the manifold, a large class of constrained optimization problems can be viewed as unconstrained optimization problems on the manifold \cite{Hu2020}.

 \subsection{COLLECTION OF MANIFOLDS}

Optimization on manifolds is a versatile framework for continuous optimization. It encompasses optimization over vectors and matrices and adds the possibility to optimize over curved spaces to handle constraints and symmetries such as orthonormality, low rank, positivity, and invariance under group actions.

One of the most common manifolds is the {\bf \gls{ccm}}, in which all the elements of the optimization variable should have a unit modulus. This is usually the case of \gls{ris} phase shift optimization problems under passive operation mode.
Hence, the \gls{mo} framework is well-suited for the \gls{ris} problems.  
Table \ref{tab:manifolds} summarizes the common real and complex types of manifolds, with particular emphasis on the complex circle manifold, also known as the  “complex one-manifold”.

A {\it complex manifold} is a manifold with a structure that locally resembles complex Euclidean space, {\it i.e.}, $\mathbb{C}^n$. This means a neighborhood is homeomorphic around every point to an open subset of $\mathbb{C}^n$. In particular, Table \ref{tab:ComplexCircle} shows the main features and applications of the Complex Circle $(\mathcal{S}^1)$ manifold.

Notice that the complex circle $\mathcal{S}^1$ is a crucial introductory example of manifolds and complex manifolds, offering valuable insights into higher-dimensional and more complex spaces used in various mathematical and physical applications. Huge practical applications deploy \gls{mo} in real-life communications systems. In the sequel, we present five classes of \gls{ris}-aided \gls{m-mimo} system applications involving \gls{mo}.

\begin{table*}[!htbp]
\centering
 \caption{Common collection of manifolds \cite{Boumal_book_2023}}
\label{tab:manifolds}
\begin{tabular}{|p{1.7cm}|p{15cm}|} 
\hline
\bf Manifolds & \bf Feature \\
\hline\hline
\bf Euclidean Space $\mathbb{R}^n$ & $ \mathbb{R}^n $ is the most straightforward example of a manifold, where each point has a local neighborhood that looks exactly like $ \mathbb{R}^n$. Flat, infinite extent, commonly used in most basic analyses. \\
\hline
\bf Circle ($\mathcal{S}^1$) & \( \mathcal{S}^1 \) represents a one-dimensional manifold (1-manifold), which can be thought of as points equidistant from a center point in 2D space, like the perimeter of a circle; intrinsic periodicity (models cyclical phenomena). \\
\hline
\textbf{Sphere ($\mathcal{S}^n$)} & $\mathcal{S}^n$ generalizes the concept of a circle and sphere to “$n$” dimensions;  e.g., \( S^2 \) is the 2D surface of a 3D ball. Compact, without boundary, intrinsic higher-dimensional analogs. {\bf Use Cases:} Modeling surfaces like Earth's surface $(S^2)$. \\
\hline
\bf Torus $(\mathcal{T}^2)$ & The 2D torus is a surface shaped like a donut, which can be defined as $\mathcal{S}^1 \times \mathcal{S}^1$, the product of two circles or, generalizing, a product of $n$ circles, closed and compact. {\bf Use Cases:} Modeling periodic boundary conditions, complex cyclical phenomena \\
\hline
{\bf Projective Space} $\mathbb{RP}^n$ & Space of lines through the origin, compact, involves projective transformation. {\bf Use Cases:} Computer vision, robotics, projective geometry. \\
\hline
\textbf{Hyperbolic Space} $\mathbb{H}^n$ & Non-Euclidean, negatively curved. {\bf Use Cases:} Representing hierarchical tree structures, complex networks. \\
\hline
{\bf Hyperplanes} & These are generalizations of planes in higher dimensions. \\
\hline
\textbf{Lie Groups:} & Smooth manifold that is also a group, with applications in physics and engineering. \textit{Examples:} \(SO(3)\), \(SU(2)\). {\bf Use Cases:} Robotics, control theory, representation of symmetries. \\
\hline
\textbf{Grassmannian} ($G(k, n)$) & Space of all \( k \)-dimensional subspaces of an n-dimensional vector space. {\bf Use Cases:} Signal processing, principal component analysis in higher dimensions. \\
\hline
\textbf{Stiefel Manifold} ($V(k, n)$): & Space of all orthonormal \( k \)-frames in \( n \)-space. {\bf Use Cases:} Multivariate statistics, optimization on orthonormal matrices. \\
\hline
\textbf{Kähler Manifold:} & A complex manifold with a Hermitian metric, deeply tied to complex and symplectic geometry. \textbf{Use Cases:} Theoretical physics, string theory. \\
\hline
\textbf{Calabi-Yau Manifold:} & A special type of Kähler manifold with a Ricci-flat metric. {\bf Use Cases:} String theory, particularly compactification methods. \\
\hline
\multicolumn{2}{|c|}{\textit{\textbf{Complex Manifolds}}} \\
\hline
\vspace{1.3mm}
\bf Complex Circle $(\mathcal{S}^1)$ & or {\it Complex 1-Manifold}: identified with the complex circle; defined as the set of all complex numbers of the unit norm, defined as: $\mathcal{S}^1 = \{ z \in \mathbb{C} \mid |z| = 1 \}$. Here, $|z|$ denotes the modulus of the complex number $z$. Manifolds modeled on complex numbers, allowing holomorphic coordinates. {\bf Use Cases:} Complex dynamics, algebraic geometry. \\
\hline
\end{tabular}
\end{table*}
\begin{table}[!ht]
\centering
\caption{Features and applications for the complex circle $(\mathcal{S}^1)$ manifold}
\label{tab:ComplexCircle}
\begin{tabular}{|p{1.4cm}|p{6cm}|} 
\hline
\bf  Feature  & \bf  Description\\
\hline\hline
\textbf{1-Dim.} & While being embedded in $\mathbb{C}$ (which is like $\mathbb{R}^2$), the complex circle $\mathcal{S}^1$ is a 1-dimensional manifold.
\\
\hline
\textbf{Compactness} & It is a closed and bounded subset of $\mathbb{C}$.
\\
\hline
\textbf{Local Structure} & Locally, around any point on $\mathcal{S}^1$, it resembles the real line $\mathbb{R}$, meaning it can be mapped one-to-one onto an open interval of $\mathbb{R}$
\\
\hline
\bf Visualization &  One can visualize $\mathcal{S}^1$ as the unit circle in the complex plane, where each point on the circle is defined by a complex number $z$ with $|z| = 1$. This can be parameterized as $z = e^{i\theta}$ for $\theta \in [0, 2\pi)$, capturing its circular nature.
\\
\hline
\bf Complex Structure &  Looking at local coordinates as a complex manifold using complex logarithms and exponential. These give the local diffeomorphisms needed to open up parts of $\mathbb{C}$. The manifold structure is given by charts that map intervals around each point to the Euclidean space $ \mathbb{C}$. 
\\
\hline
\multicolumn{2}{|c|}{\textsc{Applications of the Complex Circle Manifold}}\\
\hline
\textbf{Topology}& Understanding the structure and properties of $\mathcal{S}^1$ is fundamental in algebraic topology, which contributes to studying fundamental groups and covering spaces. 
\\
\hline
\bf Complex Structure &
$\mathcal{S}^1$ is the natural domain for periodic functions and is central in studying Fourier analysis. 
\\
\hline
 \textbf{Physics}&  The complex circle appears in various physical theories, including quantum mechanics and wave mechanics, describing spaces of phases.
 \\
\hline
\end{tabular}
\end{table}


\section{MANIFOLD LEARNING METHODS AND STRUCTURES}\label{sec:III}

Specific problems within wireless communication scenarios possess unique domains, characteristics data, and different properties. To effectively capture the essential structure of the data and optimize the performance of the wireless system, it is crucial to thoroughly understand the problem. It is important to note that the associated manifold can vary significantly from one problem to another, depending on the inherent characteristics of each situation and requirements.

By appropriately understanding the associated manifold, one can effectively leverage manifold learning techniques to transform high-dimensional data into a more manageable form, leading to better optimization in wireless systems. In the context of practical wireless system problems, manifold techniques and optimization can be employed to effectively model complex, high-dimensional spaces and enhance performance in various tasks such as signal processing, resource allocation, and network management. 

\subsection{STEPS FOR IDENTIFY A MANIFOLD}

\vspace{2mm}

\begin{enumerate}
    \item[{\bf {1)}}] {\it Identifies the Problem Domain and Requirements}: Common problems in \gls{5g}/\gls{b5g} include beamforming, interference management, user scheduling, and power control. Therefore, identifying the specific requirements and constraints such as latency, throughput, \gls{ee}, and \gls{qos}, is essential.

    \item[{\bf {2)}}] {\it Analyze the Data}:  Collect and analyze the data relevant to the optimization problem, including signal measurements, user mobility patterns, \gls{csi}, and network topology.

\item[{\bf {3)}}] {\it Understand the Dimensionality}: Determine the intrinsic dimensionality of the data. High-dimensional datasets often have a lower-dimensional structure that can be exploited. Use \gls{pca} or exploratory data analysis to estimate the true dimensionality of the data.

\item[{\bf {4)}}] {\it Identify an Appropriate Manifold Learning Technique}\cite{meila2023manifoldlearning}:
\begin{itemize}
\item[\dm]\textbf{Principal Component Analysis (\gls{pca}):}. 
This linear technique reduces dimensionality while retaining the maximum variance in the data.
\textit{Use Cases:} Effective for datasets where the important variance is linear and global structure is more important.

\item[\dm] \textbf{\gls{mds}:} Can be either linear or non-linear, and aims to preserve pairwise distances. \textit{Use Cases:} Good for visualizing distances or dissimilarities among data points.

\item[\dm] {\bf \gls{isomap}:} Non-linear and preserves global geodesic distances, useful when the data lies on a curved surface; suitable for data on a nonlinear manifold. \textit{Use Cases:} Data where the intrinsic geometry is best captured by a global isometry.

\item[\dm] \textbf{\gls{lle}:} Effective for capturing local neighborhood information, useful in highly curved manifolds. It preserves local neighborhood structure using linear reconstructions. \textit{Use Cases:} Capturing local manifold structure, suitable for highly curved manifolds.

\item[\dm] \textbf{\gls{t-sne}:} Often used to visualize high-dimensional data visualization, capturing local similarities. Non-linear and focuses on preserving local similarities. \textit{Use Cases:} Visualization of high-dimensional datasets, often used in exploratory data analysis.

\item[\dm] \textbf{\gls{umap}:}   
Non-linear, faster, and more scalable than \gls{t-sne}, preserves local and global structures. \textit{Use Cases:} Similar use cases as \gls{t-sne} but with better scalability and speed.

\item[\dm] \textbf{Laplacian Eigenmaps:} Non-linear, relies on graph-based representation, preserves local neighborhood information. \textit{Use Cases:} Data where local connectivity and local geometrical features are important.

\item[\dm] \textbf{Autoencoders:}  Non-linear, based on neural networks, encodes data into latent spaces. \textit{Use Cases:} Data with complex non-linear structures, can be used for both unsupervised and supervised learning.

\item[\dm] \textbf{Hessian Eigenmaps:}  Non-linear, focuses on preserving second-order structure (curvatures). \textit{Use Cases:} Manifolds where curvature information is crucial.

\item[\dm] \textbf{Diffusion Maps:} Non-linear, uses diffusion processes to find meaningful geometric descriptions. \textit{Use Cases:} Clustering, spectral embedding, data denoising.

    \end{itemize}

    \item[{\bf {5)}}] {\it Implement and Validate}: Implement the chosen manifold learning technique using frameworks such as \href{https://scikit-learn.org/stable}{\texttt{scikit-learn}} or custom-built solutions. Thus, validating the manifold representation by examining how well it captures the critical features of the data and supports the optimization objectives, can be done. Use cross-validation or other validation techniques to ensure the manifold model generalizes well to new data.

    \item[{\bf {6)}}] {\it Apply for Optimization}: Once the manifold is identified and validated, use it to transform and simplify the optimization problem, accordingly: for example, \textbf{Beamforming} (manifold learning can help reduce the complexity of beam selection and improve beam alignment); \textbf{Resource Allocation} (utilize manifold structures to identify clusters of users or resources to optimize allocation strategies); \textbf{Interference Management} (capture interference patterns' spatial and temporal characteristics on lower-dimensional manifolds).

\end{enumerate}


\section{METHODOLOGY FOR FORMULATING AND SOLVING OPTIMIZATION PROBLEMS WITH MANIFOLDS}\label{sec:IV}

In the following, we illustrate the \gls{mo} methodology by detailing the steps involved in addressing a real-world problem in communication systems. Specifically, we will focus on optimizing phase shifts in a \gls{ris} using manifold optimization techniques. This approach serves as an example, but the methodology can be applied to any problem involving non-convex constraints that can be represented as a manifold.
\gls{ris} is conceived to modify the propagation environment to create constructive interference patterns dynamically, thus, in practice enhancing signal strength at the \glspl{ue}, e.g., \gls{iot} devices, can be very interesting. The phase shifts introduced by \gls{ris} elements and the configurations of the \gls{m-mimo} antenna arrays constitute high-dimensional and non-linear spaces. \gls{mo} provides a framework for handling these complex optimization problems. 

\subsection*{STEP 1: PROBLEM FORMULATION}

\begin{itemize}
\item[\dm]\textbf{Define the Objective}: Clearly define the optimization problem in terms of an \gls{of}. This could be maximizing some \gls{kpi}, such as \gls{se}, \gls{ee}, or \gls{snr}. The \gls{of} is represented by $f(\boldsymbol{\theta},\boldsymbol{h})$, 
    where $\boldsymbol{h}$ and $\boldsymbol{\theta}$ represent the channel gain and the phase shift imposed by the \gls{ris}, equipped $N$ elements, respectively, with $[\boldsymbol{\theta}]_n = \alpha_n e^{j\theta_n}$.

    \vspace{2mm}
    
    \item[\dm]\textbf{Constraints}: Identify the constraints of the problem, such as power limitations, \gls{ris} phase shift constraints, and \gls{csi} requirements. In our example, 
    physically, each element in the passive \gls{ris} must satisfy the following property \(|\alpha_n e^{j\theta_n}| = 1\).
\end{itemize}

\subsection*{STEP 2: IDENTIFY THE MANIFOLD STRUCTURE}

\begin{itemize}
\item[\dm]\textbf{Manifold Description}: Establish the geometric structure of the constraint. For instance, \gls{ris} phase shifts lie on \gls{ccm}, which can be treated as a manifold \footnote{Common manifolds in wireless communications include the Stiefel manifold (for orthonormal matrices) and the Grassmann manifold (subspaces). For a complete list of Manifolds, see Table \ref{tab:manifolds}.} and described as
    \begin{equation}
        \mathcal{S}^1 = \{e^{j\theta_n} \in \mathbb{C} \mid \theta \in [0, 2\pi)\}, \quad \forall n=1,\dots,N,
    \end{equation}
    where $N$ is the total number of elements of \gls{ris}.
\end{itemize}

\subsection*{STEP 3: REFORMULATE THE OPTIMIZATION PROBLEM}

\begin{itemize}
    \item[\dm]\textbf{Manifold Representation}: Reformulate the constraint problem regarding manifold constraints. For example, if the optimization involves unit-modulus constraints, e.g., \gls{ris} phase shifts $\theta \in [0, 2\pi)$, represent these in terms of the complex exponential form $e^{j\theta}$. In our specific example, the optimization problem becomes:
    \begin{equation}
        \max_{\boldsymbol{\theta} \in \mathcal{S}^1} f(\boldsymbol{\theta},\boldsymbol{h}),
    \end{equation}
    where, $\boldsymbol{\theta}$ is a vector of phase shifts, and each element must satisfy the unit-modulus constraint.

    \vspace{2mm}
    
    \item[\dm]\textbf{Alternative Parameterization}: Use appropriate parameterizations to represent elements on the manifold in a computationally friendly way.
\end{itemize}

\subsection*{STEP 4: DEVELOP AN OPTIMIZATION ALGORITHM}

\begin{itemize}
    \item[\dm]\textbf{Initialization}: Start with an initial feasible point on the manifold. This might involve random initialization or a heuristic-based initialization.
    \begin{equation}
        \boldsymbol{\theta}^{(0)} = [\theta_1^{(0)}, \theta_2^{(0)}, \ldots, \theta_N^{(0)}]^T.
    \end{equation}

    \vspace{2mm}
     
    \item[\dm]\textbf{Gradient Descent}: Use some {\it manifold-based optimization technique} (see subsection \ref{subsec:gradient} of section \ref{sec:V}) to iteratively update the phase shifts and move towards the local optimum. The steps should ensure the updated points lie on the manifold.
    \begin{equation}
        \boldsymbol{\theta}^{(k+1)} = \boldsymbol{\theta}^{(k)} + \alpha \nabla_{\mathcal{M}} f(\boldsymbol{\theta}^{(k)}),
    \end{equation}
    where \(\alpha\) is the step size and \(\nabla_{\mathcal{M}} f(\boldsymbol{\theta}^{(k)})\) is the Riemannian gradient at the $k$-th iteration.

    \vspace{2mm}

    \item[\dm]{\bf Returning to the Manifold}: After updating the phase shifts, the updated point should be on the manifold surfaces, therefore, the retraction operator should be applied. Specifically, for the \gls{ccm} manifold, the retraction operator is given as

    \begin{equation}
        \operatorname{Retr}_{\mathcal{S}^1}(\boldsymbol{\theta}^{(k+1)}) = \frac{\left[\boldsymbol{\theta}^{(k+1)}\right]_n}{\left|\left[\boldsymbol{\theta}^{(k+1)}\right]_n \right|}, \quad \forall n={1,2,\dots,N}.
    \end{equation}
\end{itemize}

\subsection*{STEP 5: ITERATIVE OPTIMIZATION}

\begin{itemize}
    \item[\dm]\textbf{Iterative Process}: Iterate the optimization process until convergence. The stopping criterion could be based on the change in the \gls{of} value or the gradient norm at the $k$-th iteration till a small positive threshold $\epsilon$.
\end{itemize}
\begin{equation}
    \|\nabla_M f(\boldsymbol{\theta}^{(k)})\| < \epsilon.
\end{equation}

\subsection*{STEP 6: VALIDATION AND PERFORMANCE EVALUATION}

\vspace{2mm}

\begin{itemize}
    \item[\dm]\textbf{Validation}: Validate the optimized phase shifts by evaluating the reflected \gls{ris} signal's energy. Compare the performance with traditional optimization methods to demonstrate the efficiency and effectiveness of the \gls{mo}-based approach.
\end{itemize}

\section{PRACTICAL EXAMPLE AND GRADIENT DESCENT-BASED ALGORITHMS} \label{sec:V}

In this section, we elucidate a practical example of \gls{ris}-assisted solved by \gls{mo} in wireless communication systems.  In the following, we exhibit a list of gradient descend-based methods, which can be generalized to the Riemannian space and utilized in \gls{mo} strategy. 

\subsection{EXAMPLE:  BS PRECODING AND RIS PHASE SHIFT OPTIMIZATION}

Consider optimizing the beamforming weights at \gls{bs} and \gls{ris} phase shifts to maximize the sum of \glspl{sinr}. Following the steps elucidated in Section \ref{sec:III}, we can effectively solve the precoding and \gls{ris} phase shift optimization problem using \gls{mo} techniques. The key advantage of this approach is its ability to handle non-convex constraints naturally, leading to more efficient and effective optimization than more conventional methods. We start, by formulating our {\bf objective}, which is to optimize the active beamforming at the \gls{bs} (equipped with $M$ antennas) and the passive reflective beamforming at the \gls{ris} to maximize the \gls{sinr} of $K$ \glspl{ue}. This joint problem can be formulated as:
\begin{subequations}
    \begin{align}
    &\max_{\boldsymbol{W}, \boldsymbol{\theta}} \quad  f(\boldsymbol{\theta},\boldsymbol{W}) \triangleq \sum_{k=1}^{K} \frac{|\boldsymbol{w}_k^H \boldsymbol{h}_k(\boldsymbol{\theta})|^2}{\sum_{j \neq k} |\boldsymbol{w}_j^H \boldsymbol{h}_k(\boldsymbol{\theta})|^2 + \sigma^2}, 
    \\
    &\text{subject to} \quad  \boldsymbol{\theta} \in \mathcal{S}^1,
    \label{eq:constraint1}
    \\ & \phantom{\text{subject to}} \quad  \operatorname{tr}\left(\boldsymbol{W}^H \boldsymbol{W}\right) \leq P_{\max},  \label{eq:constraint2}
\end{align}
\end{subequations}
where $\sigma^2$ is the noise power.
The {\bf constraints} for the adopted problem are related to the \gls{ris} phase shifts $\boldsymbol{\theta}$, as shown in Eq. \eqref{eq:constraint1}, which should lie on the unit complex circle $\left(|e^{j\theta_i}| = 1, \, \forall i \right)$, and about the active beamforming vector $\boldsymbol{W}$, which should satisfy power norms, i.e., it must not exceed the available power budget at the \gls{bs} ($P_{\max}$), as described in Eq. \eqref{eq:constraint2}.

The optimization problem above can be addressed using \gls{mo}. By treating the beamforming weights and \gls{ris} phase shifts as optimization variables on a product manifold of spheres and circles, we identify the manifold structure, therefore, the active beamforming matrix $\boldsymbol{W}$ lies on a Stiefel manifold (for orthonormal matrices), and the phase shifts $\boldsymbol{\theta}$ lie on a \gls{ccm} $\mathcal{S}^1$. These manifolds can be described as:  
\begin{align}
    \mathcal{M}_W &= \{\boldsymbol{W} \in \mathbb{C}^{M \times K} \mid \boldsymbol{W}^H \boldsymbol{W} = \mathbf{I}_K\},
    \\
    \mathcal{M}_\theta &= \{e^{j\theta} \in \mathbb{C} \mid \boldsymbol{\theta} \in [0, 2\pi)\}.  
\end{align}
 
With the given step, we can reformulate the optimization problem in terms of the manifold constraints. The optimization problem becomes:  
\begin{equation}
    \max_{\boldsymbol{W} \in \mathcal{M}_W,  \boldsymbol{\theta} \in \mathcal{M}_\theta, \boldsymbol{p}} f(\boldsymbol{W}, \boldsymbol{\theta}),
\end{equation}
where $\boldsymbol{p}$ is the power allocation corresponding to the \glspl{ue}. Notice that $\boldsymbol{p}$ emerges from the assumption, of $\boldsymbol{W} \in \mathcal{M}_{\boldsymbol{W}}$, and its optimization is very consolidated in the literature.

The {\bf initialization} of the algorithm can start with an initial feasible point on the manifold. This could be a random initialization or based on a heuristic: 
\begin{equation}
    \boldsymbol{W}^{(0)}, \boldsymbol{\theta}^{(0)}.
\end{equation}

Thus, the Riemannian gradient descent to iteratively update the beamforming matrix and phase shifts. The update rule involves computing the Riemannian gradient and ensuring the updated points lie on the manifold.  
\begin{equation}
    \boldsymbol{W}^{(k+1)} = \boldsymbol{W}^{(k)} + \alpha_W \nabla_{\mathcal{M}_W} f(\boldsymbol{W}^{(k)}, \boldsymbol{\theta}^{(k)}),
\end{equation}
\begin{equation}
    \boldsymbol{\theta}^{(k+1)} = \boldsymbol{\theta}^{(k)} + \alpha_\theta \nabla_{\mathcal{M}_\theta} f(\boldsymbol{W}^{(k)}, \boldsymbol{\theta}^{(k)}), 
\end{equation} 
where $\alpha_W$ and $\alpha_\theta$ are the step sizes, and $\nabla_{\mathcal{M}_W} f$ and $\nabla_{\mathcal{M}_\theta} f$ are the Riemannian gradients.  

The updated values should return to the manifold surfaces, and this can be done by applying the retraction operator

\begin{align}
    &\boldsymbol{W}^{(k+1)} = \operatorname{Retr}\left( \boldsymbol{W}^{(k+1)} \right),
    \\
    &\boldsymbol{\theta}^{(k+1)} = \operatorname{Retr}\left( \boldsymbol{\theta}^{(k+1)} \right).
\end{align}

Finally, we can {\bf iterate} the optimization process until convergence. The stopping criterion could be based on the change in the \gls{of} value or the gradient norm.  
\begin{align}
    & \|\nabla_{\mathcal{M}_W} f(\mathbf{W}^{(k)}, \boldsymbol{\theta}^{(k)})\| < \epsilon, 
    \\
    &\|\nabla_{\mathcal{M}_\theta} f(\mathbf{W}^{(k)}, \boldsymbol{\theta}^{(k)})\| < \epsilon,
\end{align}
where $\epsilon$ is a small positive threshold.  

Thus, we can validate the optimized beamforming matrix and phase shifts by evaluating the \glspl{sinr}. We should compare the performance with traditional optimization methods to demonstrate the efficiency and effectiveness of the manifold-based approach.  

\subsection{GRADIENT DESCENT-BASED ALGORITHMS} \label{subsec:gradient}

Manifold-based optimization techniques include \gls{rn}, \gls{rcg}, \gls{rtr}, and \gls{rgd}. These techniques are powerful for optimizing functions constrained to manifolds, taking into account the non-Euclidean geometry of the optimization space. By leveraging the geometric properties of manifolds, these methods enable efficient and effective optimization. This is particularly beneficial in complex wireless communication systems, such as \gls{ris}-enabled \gls{m-mimo}, where gradients are computed directly on the manifold, navigating over the curvature of the space. This approach ensures efficient convergence to locally optimal solutions.

\vspace{2mm}

\noindent{\em \gls{rgd}}: The \gls{rgd} is an optimization technique that extends the classical gradient descent method to Riemannian manifolds. The optimization process is constrained to a curved space rather than a flat Euclidean space. The key idea is to iteratively move towards the local optimum while ensuring that each update remains on the manifold. The \gls{rgd} Algorithm steps are as follows:

\begin{enumerate}
    \item[\dm]\textbf{Initialization:} Start with an initial feasible point on the manifold.
    
    \item[\dm]\textbf{Gradient Computation:} Compute the Riemannian gradient, which is the projection of the Euclidean gradient onto the tangent space of the manifold.
    
    \item[\dm]\textbf{Update Rule:} Move in the direction of the Riemannian gradient by a step size, ensuring the updated point lies on the manifold. This often involves a retraction operation that maps the point back onto the manifold.
    
    \item[\dm]\textbf{Iteration:} Repeat the gradient computation and update steps until convergence.
\end{enumerate}

\begin{algorithm}
\label{alg:rgd}
\small
\caption{\gls{rgd} Algorithm}
\begin{algorithmic}[1]

\STATE \textbf{Input:} Initial point $\mathbf{x}_0$ on manifold $\mathcal{M}$, step size $\alpha$, max iterations $K$

\STATE \textbf{Output:} Optimized point $\mathbf{x}^*$ on the manifold
\algorithmiccomment{\it Input and output specifications}

\STATE $k \gets 0$ \algorithmiccomment{ Initialization of the iteration counter}

\WHILE{$k < K$ and not converged} 
    \STATE Compute Riemannian gradient $\operatorname{grad} f(\mathbf{x}_k)$ \algorithmiccomment{\it Project Euclidean gradient onto tangent space}

    \STATE Update: $\mathbf{x}_{k+1} \gets \mathcal{R}_{\mathbf{x}_k}(-\alpha \operatorname{grad} f(\mathbf{x}_k))$ 
    \algorithmiccomment{\it Utilize retraction operator to map point back to manifold} 
    
    \STATE $k \gets k + 1$
\ENDWHILE

\STATE \textbf{return} $\mathbf{x}_k$ \algorithmiccomment{\it  Return of the final optimized point}

\end{algorithmic}
\end{algorithm}

\vspace{2mm}

\noindent{\em \gls{rn}:} \gls{rn} method is an extension of the classical Newton's method to Riemannian manifolds. It uses second-order information (Hessian) to achieve faster convergence compared to the gradient descent.

\vspace{2mm}

\noindent{\em \gls{rcg}:} \gls{rcg} is an adaptation of the conjugate gradient method to Riemannian manifolds. It combines the efficiency of the conjugated gradient method with the geometric constraints of manifolds.

\vspace{2mm}

\noindent{\em \gls{rtr}:} \gls{rtr} methods extend trust-region methods to the Riemannian manifolds. These methods iteratively solve a local approximation of the optimization problem within a “trust region” around the current point.

 Utilizing \gls{mcs} or other simulation techniques is useful to evaluate the algorithm's performance under different scenarios. Based on simulation results, we can also refine the algorithm for better performance by adjusting parameters and re-evaluating. Refinement based on simulations is a critical step in the optimization process, especially when dealing with complex systems like those involving manifolds. In this context, the refinement process based on simulations might involve specific steps such as:

\begin{itemize}
    \item[\dm] \textit{Gradient and Hessian Adjustments:} Fine-tuning the computation of Riemannian gradients and Hessians to ensure they accurately capture the manifold's geometry.
    \item[\dm] \textit{Retraction Operations:} Modifying the retraction operations to better map updated points back onto the manifold.
    \item[\dm] \textit{Step Size Adaptation:} Adjusting the step size dynamically based on the manifold's curvature to ensure efficient convergence.
    \item[\dm] \textit{Constraint Handling:} Refining how additional constraints are incorporated into the optimization problem, possibly by adjusting the indicator functions or penalty terms.
\end{itemize}

The goal is to iteratively improve the algorithm or model based on empirical evidence from simulations, leading to better performance in the target application. How refinement is typically done based on simulations is summarized in Algorithm \ref{alg:refinement}. 

Some open-source libraries support manifold structures, such as  \href{https://www.manopt.org/}{\texttt{Manopt tools}}, which implements a bunch of manifold collections available on \texttt{Manopt} (Matlab) or \texttt{Pymanopt} (Python), which can be useful for obtaining solutions of many diverse problems \cite{manopt2014}, 
\cite{Boumal_book_2023}.

\begin{algorithm}
\caption{Refinement Based on Simulations}
\label{alg:refinement}
\small
\begin{algorithmic}[1]
    \STATE \textbf{Initial Simulation:}
    \STATE Run initial simulations to evaluate the performance of the current algorithm or system configuration.
    \STATE Collect performance metrics such as \gls{se}, \gls{ee}, error rates, throughput, etc.
    
    \STATE \textbf{Performance Analysis:}
    \STATE Examine the simulation results to identify strengths and weaknesses.
    \STATE Determine which aspects are underperforming or causing issues.
    
    \STATE \textbf{Parameter Adjustment:}
    \STATE Modify parameters of the algorithm or system based on the analysis.
    \STATE Make necessary changes to the algorithm itself if required.
    
    \STATE \textbf{Re-simulation:}
    \STATE Perform new simulations with the adjusted parameters or modified algorithm.
    \STATE Compare the new simulation results with the previous ones.
    
    \STATE \textbf{Iteration:}
    \STATE Repeat the process of analysis, adjustment, and re-simulation multiple times.
    \STATE Continue iterating until performance metrics converge to satisfactory levels.
    
    \STATE \textbf{Validation:}
    \STATE Use cross-validation techniques to ensure generalization to different scenarios.
    \STATE Test the refined system under various conditions for robustness.
    
    \STATE \textbf{Final Tuning:}
    \STATE Perform fine-tuning of parameters to achieve the best possible performance.
    \STATE Use advanced optimization techniques if necessary.
\end{algorithmic}
\end{algorithm}


\section{MANIFOLDS IN RIS-AIDED MASSIVE MIMO SCENARIOS: FOUR REAL APPLICATIONS} \label{sec:VI}

In the sequel, we discuss in detail four applications of \gls{mo} framework to wireless communication systems, precisely, by applying different \gls{mo} tools for \gls{ris}-aided \gls{m-mimo} systems optimization. We use specific optimization tools based on \gls{mo} procedures and methodologies, confirming the promising results attainable by applying \gls{mo} tools.



Subsection \ref{sec:app1} ({\it Application 1}) presents an efficient algorithm to maximize the minimum rate of the network. This algorithm enhances the system's rate in a maximally fair manner by leveraging \gls{rcg} within \gls{ccm}. We then compare the performance and complexity of the proposed technique against benchmarks commonly used in the literature for \gls{ris}-assisted \gls{m-mimo}, such as \gls{sca} and \gls{ao}. The proposed algorithm demonstrates significant gains over these benchmarks in terms of both performance and complexity.

Subsection \ref{sec:app2} ({\it Application 2}) discusses how effective can be an \gls{rm}-based \gls{ris} phase shift optimization method for optimizing the \gls{ris} phase shifts and the \gls{bs} combining vectors. This technique is designed to minimize the \gls{ul} transmit power of the \glspl{ue} in the context of an \gls{iot} network supported by \gls{ris}. Initially, the joint optimization for the phase shift and combining vectors is formulated aiming to improve the overall system \gls{ee} by minimizing the \gls{ul} transmit power while guaranteeing a minimum \gls{qos} for the devices. To tackle the non-convexity of the problem,  \gls{rm}-based iterative alternating optimization (i-\gls{ao}) technique was deployed. The \gls{rm} iterative alternating optimization (\gls{rm}-\gls{ao}) algorithm greatly improves the system's \gls{ee} and {\it resource efficiency}\footnote{The \gls{se} $\times$ \gls{ee} tradeoff.} when compared to the non-\gls{ris} \gls{mu} \gls{m-mimo} system. 

Subsection \ref{sec:app3} ({\it Application 3}) discusses how to apply the \gls{mo} framework to the \gls{ris} phase shifts optimization for intra-cell pilot reuse and the associated channel estimation. Relying upon the knowledge of only statistical \gls{csi},  the \gls{ris} phase shift optimization highlights the remarkable performance improvements achieved by the proposed scheme (for both \gls{ul} and \gls{dl} transmissions).

Subsection \ref{sec:app4} ({\it Application 4})  presents an optimization problem in a \gls{ris}-based \gls{gf} random access (\gls{gf}-\gls{ra}) protocol, formulated to obtain a uniform high-gain reflected multi-beam in the intended direction while keeping low reflection gains in the unintended ones, subject to the unit modulus
constraint. A \gls{mo} framework reveals to be a promising tool to solve such optimization
problems.

\subsection{\textbf{Application 1: Maximizing Fairness in RIS-Aided m-MIMO DL}} \label{sec:app1}

To achieve efficient wireless communication, it is crucial to utilize available resources effectively, particularly in the context of future wireless networks. With the rise of the passive \gls{ris}, another variable should be appropriately optimized, the passive reflective beamforming. Jointly, the active precoding at the \gls{bs} must not be overlooked. By leveraging advanced techniques, such as \gls{zf} precoding, we can achieve a condition where the total interference of the system can be completely nullified. 

To accomplish this efficiently, accurate \gls{csi} is crucial. By knowing the \gls{csi}, the \gls{bs} can effectively manage the system, by sending specific power for the \glspl{ue} and configuring the \gls{ris}, to enhance the overall performance of the network. Besides, fairness in future wireless communications is fundamental since it can provide access for all \glspl{ue} in the cell from an equality perspective. With the \gls{ris} assistance, there is more room for improving fairness, demanding thus research efforts to develop techniques to achieve such a challenging target.


We aim to fairly maximize the \glspl{ue}’ communication performance. To achieve this, we address the max-min fairness problem, focusing on maximizing the common \gls{sinr} across all \glspl{ue}. To this end, the reflective passive beamforming $\boldsymbol{\theta}$ at the \gls{ris} should be optimized, subject to the passive \gls{ris} constraints. We demonstrate that this problem can be formulated as a {\it minimization of inverse-summation of eigenvalues} of channel matrix under the complex circle manifold. Consequently, optimization tools such as manifold techniques, specifically, the \gls{rcg}, can be appropriate for solving this problem, ensuring that the obtained solution remains within the manifold formed by the non-convex constraint.

\noindent
{\bf System Model and Problem Formulation}:  Consider a \gls{mu} \gls{m-mimo} system operating in \gls{dl} mode, with assistance of a \gls{ris} composed of $N$ reflecting elements, where $K$ single-antenna devices are served simultaneously by a \gls{bs} equipped with $M$ antennas, as illustrated in Figure \ref{fig:subfig_w1}. 

To solve the intended problem, we need to optimize the active beamforming at the \gls{bs} ($\boldsymbol{W}$) and the passive reflective beamforming at the \gls{ris} ($\boldsymbol{\theta}$). This problem involves two main constraints:

\begin{itemize}
\item[$\bullet$] Ensuring that the optimized active beamforming at the \gls{bs} consumes no more than the available power budget  ($P_{\max}$). 
\item[$\bullet$] Ensuring that the optimized reflective passive beamforming obeys the passive constraint of the \gls{ris}, i.e., all elements should maintain the unit-modulus.
\end{itemize}

Some studies in the literature have already addressed the max-min problem, focusing on optimizing both active and passive beamforming. Solutions based on diverse techniques have been proposed, such as \gls{sdr} with \gls{sca} \cite{9246254}, penalty method followed by \gls{sca} \cite{9133118}, and even \gls{fp} with \gls{sca} \cite{9087848}. It is noteworthy that all these techniques leverage from the \gls{ao}, where this methodology is iterative and optimizes each variable of interest sequentially while keeping the others constant, and is repeated until convergence. Although all these works proposed different ways to convex the original non-convex max-min problem, proposing interesting sub-optimal solutions, their performance in real-world scenarios can be harmed due to their complexity. The main factor is that they suffer from high running time since they rely on \texttt{CVX}. Due to the interior point method implemented therein, this may lead to relatively high computational complexity. To further reduce the complexity and, consequently, the time processing, in the following, we propose a method of optimizing the minimum rate in a {\bf two-step} process, which can considerably reduce the complexity and provide interesting performance gain, as we will see further ahead.

\begin{figure}[!htbp]

\begin{subfigure}[b]{1\linewidth}
\centering 
\includegraphics[width=1\linewidth]{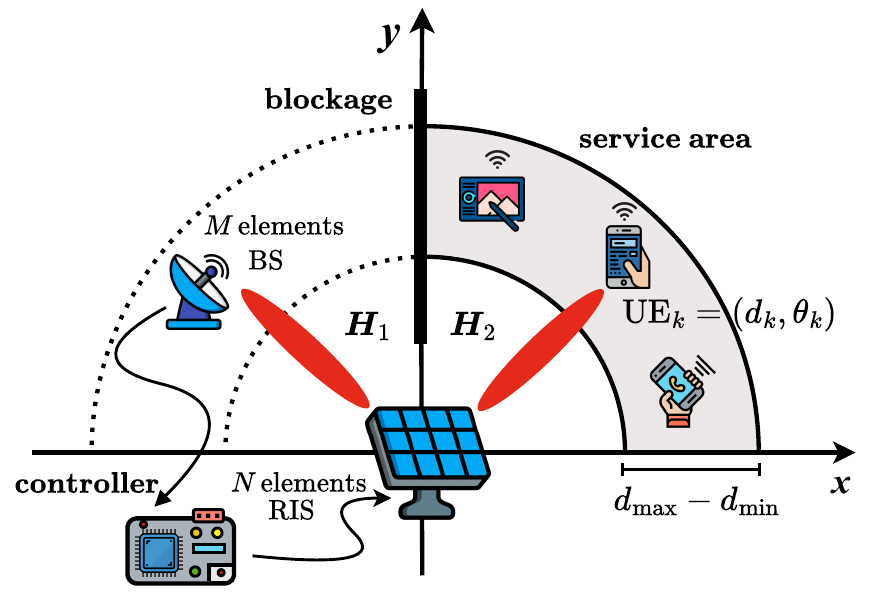}
\caption{ System model of a \gls{dl} \gls{ris}-aided \gls{mu} \gls{m-mimo} system.}
\label{fig:subfig_w1}
\end{subfigure}

\begin{subfigure}[b]{1\linewidth}
\centering
\includegraphics[width=1\linewidth]{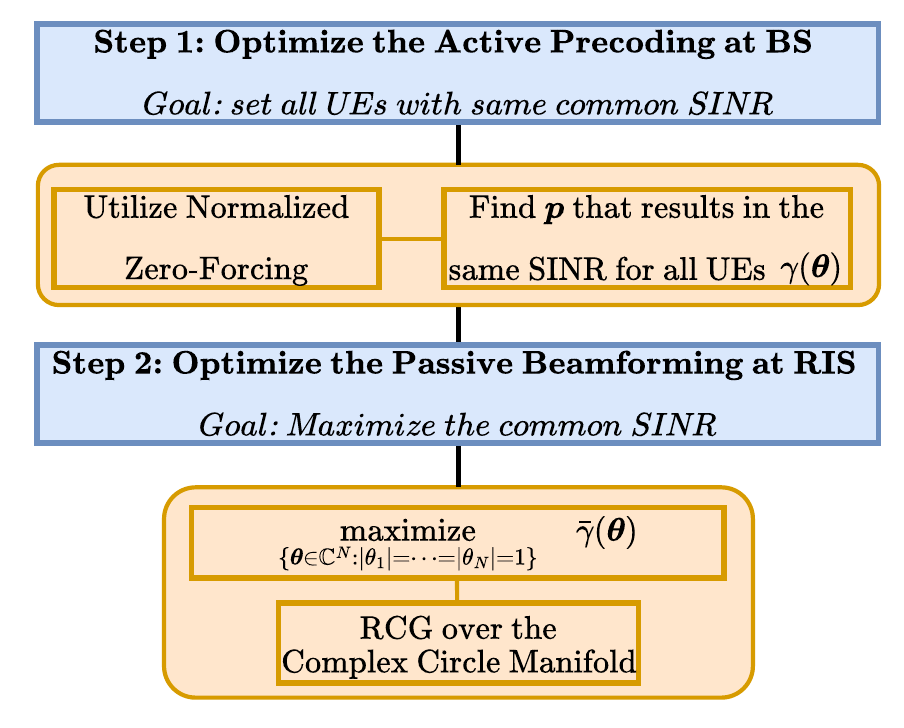}
\caption{Steps for the proposed algorithm.}
\label{fig:subfig_w2}
\end{subfigure}

\begin{subfigure}[b]{1\linewidth}
\centering
\includegraphics[width=1\linewidth]{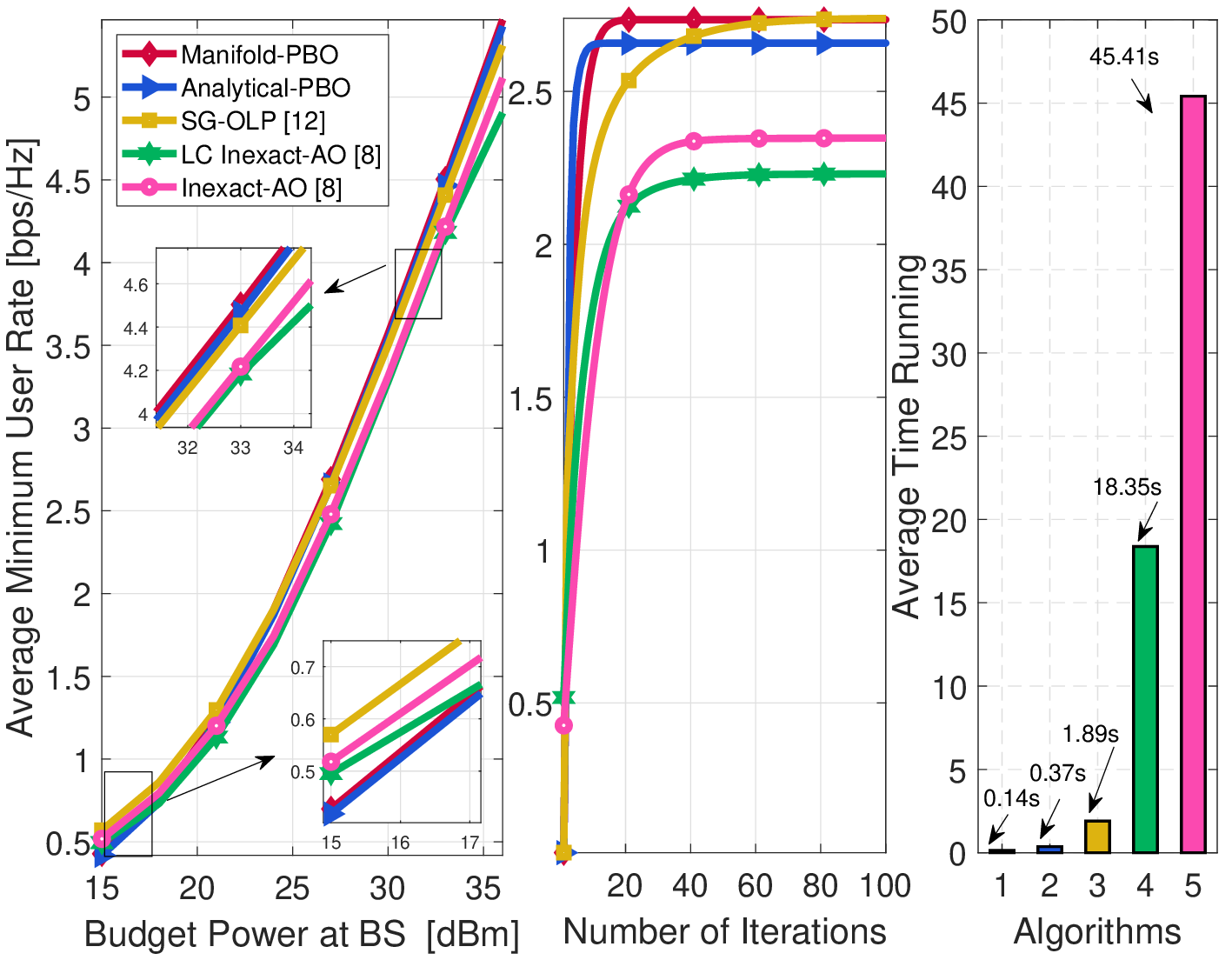}
\caption{
\hspace{2mm}
\textit{\textbf{i}}. Avg. $\gamma$ $\times P_{\max}$ 
\hspace{4mm}
\textit{\textbf{ii}}. Avg. $\gamma$ $\times N_i$ \hspace{4mm}  
\textit{\textbf{iii}}. Avg. running time  }
\label{fig:subfig_w3}
\end{subfigure}

\caption{{\textsc{Application 1}:} \small {\bf (a)} System model of the considered network; \, {\bf (b)} Steps of the proposed algorithm; \, {(\bf c)}  Simulation results.}
\end{figure}

\vspace{2mm}
\noindent{\bf Proposed Solution} The proposed solution has as the main factor avoiding the \gls{ao} methodology, remarkably reducing the complexity. Therefore, 
to this end, the proposed effective approach to solving the max-min optimization problem consists of controlling each beamforming function separately, providing a two-step algorithm that does not rely on an iterative process. With this methodology, one can design active beamforming to provide the same \gls{sinr} (denoted as $\gamma$) for all the \glspl{ue}, which we denote as {\bf step 1} of the algorithm, while the following task, {\bf step 2}, we concentrate on designing the passive beamforming aiming to maximize the common \gls{sinr}. To be more specific: 

\vspace{2mm}
\noindent {{\bf Step 1}} -- \textit{Optimizing the Active Precoding at \gls{bs}}: 

\begin{itemize}
\item[\dm]Find active precoding that can provide the same \gls{sinr} condition among all the served \glspl{ue}. To achieve this configuration, we provide the normalized \gls{zf} precoding, eliminating the interference suffered by the \glspl{ue}. However, the normalized \gls{zf} does not guarantee the equal \gls{sinr} conditions between the \glspl{ue}; thus,  the power allocation referring to each \gls{ue} should be optimized yet, aiming to achieve the maximum fairness condition. Fortunately, the \gls{zf} precoding turns out the \glspl{sinr} of \glspl{ue} as a simple linear function of its respective power allocated  \cite{9352958}, enabling to find a matrix of allocated power that obeys the maximum power budget constraint, in simple and expeditious closed-form. 
We will elucidate how to optimize the passive beamforming at the \gls{ris}.
\end{itemize}

\noindent {{\bf Step 2}} -- \textit{Optimizing the Passive Beamforming at the \gls{ris}}: 

\begin{itemize}
    \item[\dm]The major problem is how to maximize the common \gls{sinr} concerning the reflective passive beamforming. This problem can be formulated as a {\it minimization of inverse-summation of eigenvalues}, which has an easy and direct way to obtain Euclidean derivative. Therefore, the core of step 2 is to utilize the \gls{rcg} to optimize the passive reflective beamforming in view to increase as high as possible the common rate shared between all the \glspl{ue}. 
Both steps are illustrated in Figure \ref{fig:subfig_w2}.
\end{itemize}

Figure \ref{fig:subfig_w3} compares the performance and complexity of the proposed method for maximizing the minimum rate with the following benchmark schemes: 

\begin{itemize}
\item[\dm]{\it Inexact-\gls{ao}}\cite{9246254}:  
both the active beamforming and passive reflective beamforming are updated sequentially by \gls{ao} method. Specifically, 
active beamforming was solved by \texttt{CVX} as \gls{socp}, while the passive beamforming problem was solved deploying \gls{sca}. 
To increase the min-\gls{sinr} value at each iteration until the convergence, in this approach, only need to find an update for $\boldsymbol{W}$ and $\boldsymbol{\theta}$ . 
\item[\dm]{\it Low Complexity Inexact-\gls{ao}}\cite{9246254}: This method is similar to the Inexact-\gls{ao}; however, herein, for the passive beamforming, the subgradient projection method is utilized to avoid further complexity.

\item[\dm]{\it SG-OLP} \cite{10264820}: In the \gls{sg}-\gls{olp}, the strategy is optimizing the active precoding at \gls{bs} through the \gls{olp} as proposed in \cite{10264820}. 
Because of reducing the complexity, the \gls{sg} method is utilized to optimize the passive beamforming.

\item[\dm]{\it Analytical \gls{pbo}:} In this method, we solve exactly the proposed problem, given by Fig. \ref{fig:subfig_w2}; however, for the {\bf step 2}, we derived a closed-form solution for each element of $\boldsymbol{\theta}$, and updated $\theta_n$ sequentially until the convergence.
\end{itemize}

Figure \ref{fig:subfig_w3}{.(\textit{\textbf{i}}) depicts the average common rate $\Bar{\gamma}$ (equivalent to the minimum rate) at \glspl{ue} versus the maximum transmit power budget $P_{\max}$. First, it is observed that the proposed approach considerably outperforms the three benchmark schemes over a wide range of $P_{\max
}$. This demonstrates the potential of our proposed design, i.e., to optimize the \gls{zf} precoding for achieving equal rate conditions and optimize the passive beamforming to maximize the common rate. We also can see that under a lower power regime, $P_{\max} \leq 19$ dBm, both Inexact-\gls{ao}, Low-Complexity Inexact-\gls{ao}, and \gls{sg}-\gls{olp} methods can outperform the proposed method. This is justified because the \gls{zf} solution does not operate well for the lower-power regime, differently for the high-power regime, where it is known to be optimal \cite{8741198}.

Figure \ref{fig:subfig_w3}{.(\textit{\textbf{ii}})} depicts the convergence behavior of our proposed approach, where the maximum transmit power is set as $P_{\max}=25$ dBm. It is observed that our \gls{rcg} proposed method monotonically increases the common \gls{sinr} $\Bar{\gamma}$ value over iterations, thus leading to a converged solution. We also can see that all considered algorithms converge in different values. The analytical \gls{pbo} converges fastest, and the Inexact \gls{ao} converges slowest. Although the proposed \gls{rcg} does not achieve the convergence fastest, we can see that it can achieve the best performance, with $\approx 2.8$  [bps/Hz]. The obtained gains over the Analytical \gls{pbo}, Inexact-\gls{ao}, and Low-Complexity Inexact-\gls{ao} are  $3.32\%$, $19.15\%$, and $26.13\%$, respectively. Furthermore, concerning the \gls{sg}-\gls{olp} method, both achieve about the same value in the convergence for the analyzed value of $P_{\max}$, proofing the potential of the proposed solution since the running time for the manifold approach is substantially lower, as we 
discuss in the following.

Figure \ref{fig:subfig_w3}{.(\textit{\textbf{iii}})} illustrates the average running time of each algorithm considered. From this figure, we can see how it is important to consider the \textbf{step 1} of the proposed algorithm since this step is crucial for time reduction, providing low-running time over the Inexact-\gls{ao} and Low Complexity Inexact-\gls{ao} algorithms. This is expected, as \textbf{step 1} eliminates the \gls{ao} methodology from the algorithm, remarkably reducing its complexity. The high time running for Inexact-\gls{ao} is attributed to the repeated solving of two different convex problems until convergence, which is time-consuming due to its dependence on \texttt{CVX}. On the other hand, the Low-Complexity Inexact-\gls{ao} can reduce the complexity since instead of two convex problems, it solves just one convex problem (active beamforming at \gls{bs}), while the subgradient method is utilized for passive beamforming. Concerning the \gls{sg}-\gls{olp} method, our proposed approach still can be further promising in terms of complexity since the \gls{olp} method requires solving a fixed point equation followed by the \gls{sg} method, iteratively, up to the 
convergence. On the other hand, for this problem, our proposed method does not require any iterative process over the active and passive beamforming.

\subsection{\textbf{Application 2:  RIS-Aided Energy-Efficient mMIMO for IoT Systems}}
\label{sec:app2}

The need for efficient wireless communication is more critical than ever, especially with the rise of \gls{iot} applications. Optimizing \gls{ul} power allocation is essential for enhancing communication performance. By leveraging advanced technologies such as \gls{ris} combined with the linear \gls{mmse} receiver, significant improvements can be achieved in how \gls{iot} devices communicate with \glspl{bs}. 
To do this efficiently, instantaneous \gls{csi} is crucial for optimizing communication. By knowing the \gls{csi}, the \gls{bs} can effectively allocate \gls{ul} power and control the \gls{ris} phase shifts. This information allows the \gls{bs} to send specific power indicators and phase shift vectors to \gls{iot} devices, enhancing their communication with the \gls{bs}. This application is about how to get the most energy-efficient \gls{m-mimo} systems working with \gls{ris}. One can use optimization tools based on \gls{mo} methods and procedures, confirming the suitable outcomes achieved using \gls{mo} approach. 

To tackle the non-convexity of the problem, an \gls{rm}-based \gls{iao} technique was deployed. The \gls{rm} \gls{iao} algorithm makes the system much more efficient in terms of \gls{ee} and resource-efficient (\gls{se}-\gls{ee} tradeoff) than the non-\gls{ee} \gls{mu} \gls{m-mimo} system. \gls{ee} \gls{rm}-based optimization methodology for \gls{ris}-aided \gls{m-mimo} includes formulate and solve the optimization problem using \gls{rm}-based \gls{iao} algorithm:
\begin{itemize}
\item[\dm]Formulate the \gls{ee} power minimization problem in the \gls{ul} \gls{ris}-aided \gls{m-mimo} \gls{iot} network and establish a solution methodology based on the \gls{rm} approach.
\item[\dm]Develop an optimization methodology for \gls{ee} power minimization problem in different \gls{ris}-aided \gls{m-mimo} system configurations by deploying specific \gls{mo} formulations and tools.
\item[\dm]\gls{kpi}, including: a) \gls{se} $\times$ \gls{ee} tradeoff maximization; b) precoding/combining design; c) effective power allocation strategies for \gls{m-mimo}.

\end{itemize}

We deploy an \gls{rm}-based \gls{ao} technique for optimizing the \gls{ris} phase shifts and the \gls{bs} receiver-combiner vectors. In the context of an \gls{iot} network with \gls{ris} support, this technique aims to reduce the \gls{ul} transmit power of the \glspl{ue}. Initially, the joint optimization for the \gls{ris} phase shift and \gls{bs} combining vectors is formulated to improve the overall system \gls{ee} by minimizing the \gls{ul} transmit power while guaranteeing a minimum \gls{qos} for the devices.

\vspace{2mm}
\noindent{\bf System Model}: Consider the \gls{ul} of multiple access \gls{m-mimo} systems, where $K$ \gls{iot} single-antenna devices transmit simultaneously to a \gls{bs} equipped with $M$ antennas, representing a typical \gls{m-mimo} scenario (Figure \ref{fig:subfig_d1}). Strategically positioning a \gls{ris} can substantially improve system reliability, facilitating communication between users and the \gls{bs}. The \gls{ris} contains $N$ reconfigurable reflecting elements. The \gls{ris} delivers a phase-shifted version of the transmitted signal, maximizing the composite channel gain. Moreover, the \gls{dl} dual problem can be considered similarly, being omitted herein.

\begin{figure}[!htbp]

\begin{subfigure}[b]{1\linewidth}
\centering 
\includegraphics[width=1\linewidth]{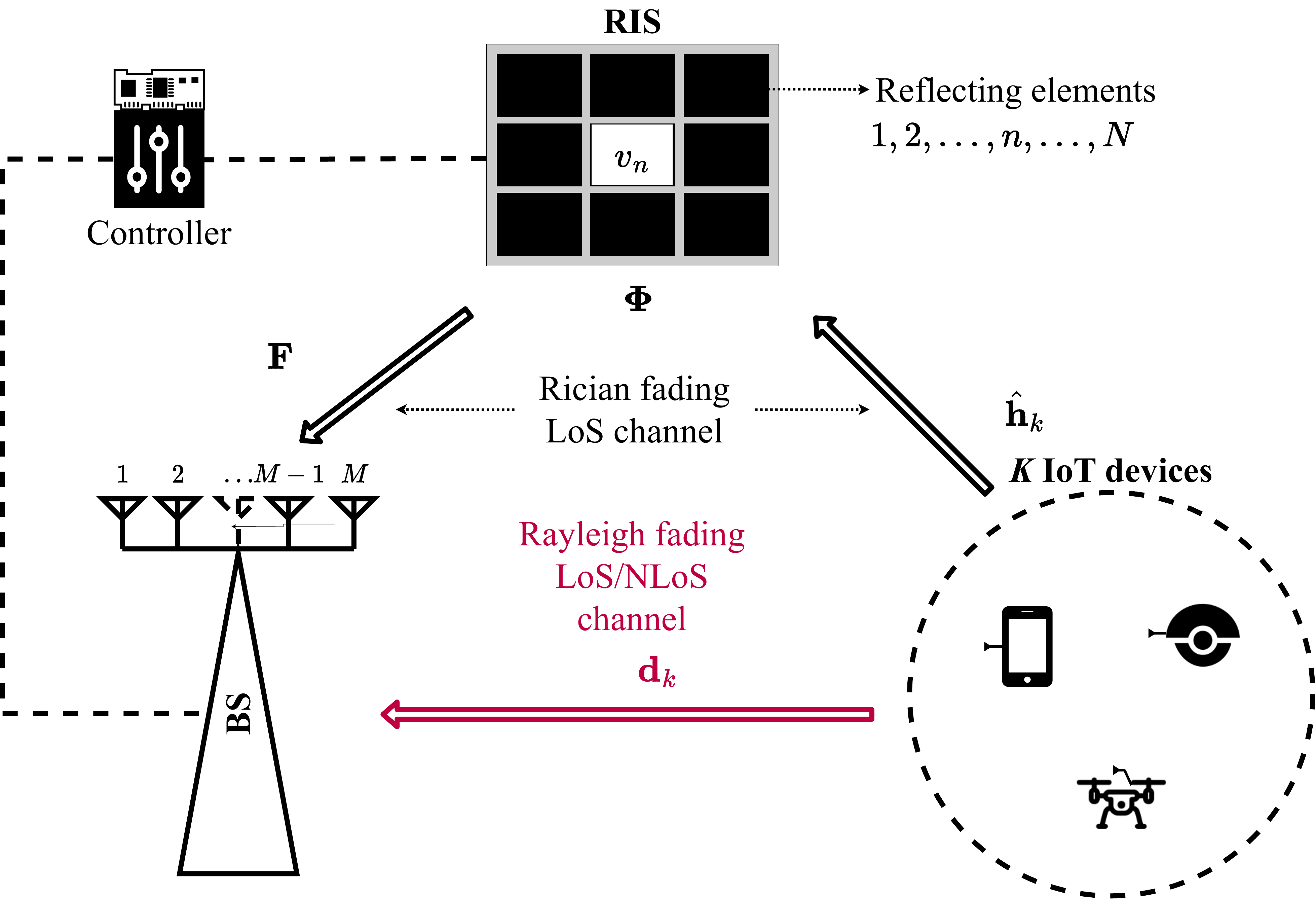}
\noindent \caption{Illustration of a passive \gls{ris}-aided \gls{mu} \gls{m-mimo} system.}
\label{fig:subfig_d1}
\end{subfigure}

\begin{subfigure}[b]{1\linewidth}
\centering 
\includegraphics[width=1\linewidth]{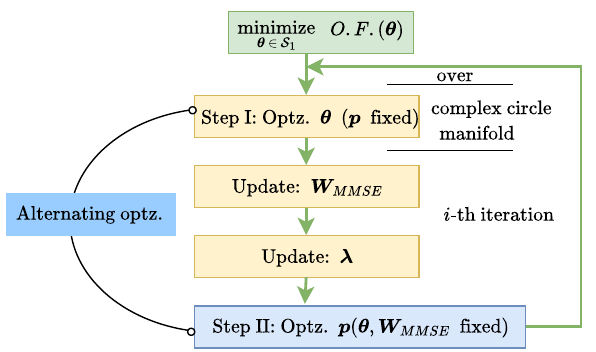}
\caption{\gls{iao} steps for the proposed algorithm attain convergence. 
}
\label{fig:subfig_d2}
\end{subfigure}

\begin{subfigure}[b]{1\linewidth}
\centering 
\includegraphics[width=1\linewidth]{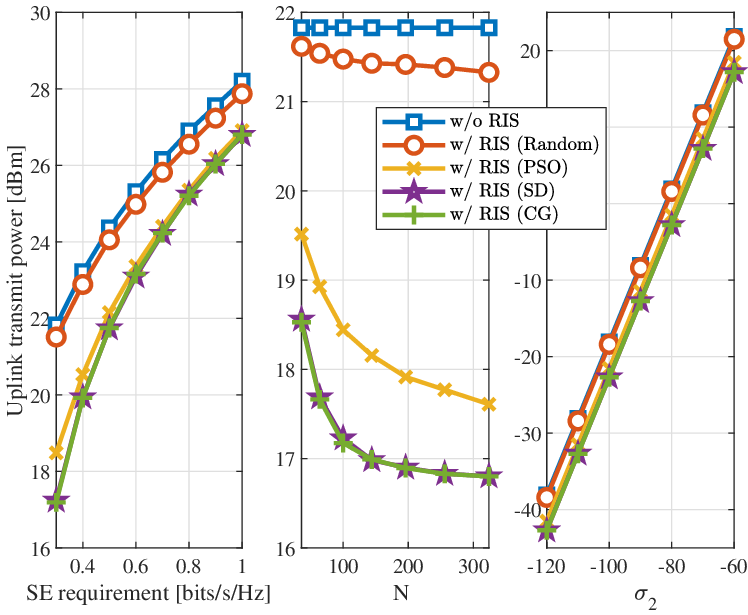}
\caption{ \phantom{aa}
\textit{\textbf{i}.} $P_{\text{UT}}$ $\times$ min. \gls{se} \hspace{6.5mm} 
\textit{\textbf{ii}.} $P_{\text{UT}}$ $\times N$ \hspace{6.5mm} 
\textit{\textbf{iii}.} $P_{\text{UT}}$ $\times \sigma^2$ \phantom{noise}}
\label{fig:subfig_d3}
\end{subfigure}

\caption{{\textsc{Application 2}:} \small  {\bf (a)} System model for the considered network;  {\bf (b)} Steps of the proposed algorithm; {\bf (c)} \gls{ul} transmit power ($P_{\text{UT}}$) with $K=8$ varying: {\it i)} min \gls{se}; {\it ii)} number of reflective elements $N$; {\it iii)} noise power at \gls{bs}.}
\label{fig:RIS_M-MIMO}
\end{figure}

The signal's power is carefully controlled when each \gls{iot} device transmits data. The data signal of an \gls{iot} device combines its transmit power and a normalized data symbol. This setup ensures that each device can communicate effectively while minimizing interference with other devices.

\vspace{2mm}
\noindent{\bf Problem Formulation}: to minimize the \gls{ul} power while maintaining high \gls{se}\footnote{The metric  \gls{se}  measures how effectively the available bandwidth is utilized. The \gls{se} of an \gls{iot} device depends on its transmit power, the combining vector at the \gls{bs}, and the channel conditions. By optimizing these factors, one is doing \gls{se} maximization, ensuring that each device transmits data efficiently.}, we need to optimize the \gls{ris} phase shift vector ($\boldsymbol{\theta}$), the \gls{bs} beamforming matrix ($\boldsymbol{W}$), and the power vector of the devices ($\boldsymbol{p}$). This involves several constraints:
\begin{itemize}
\item[$\bullet$] Ensuring each device meets its \gls{qos} requirements.
\item[$\bullet$] Maintaining the unit-modulus of the \gls{ris} phase shift.
    \item[$\bullet$] Keeping the \gls{bs} combining matrix normalized.
    \item[$\bullet$] Ensuring the device power stays within maximum limits ($P_{\text{max},k}$).   
\end{itemize}

\vspace{2mm}
\noindent{\bf Proposed Solution}: an effective approach to solving this optimization problem is the \gls{ao}. This method involves iterative and sequentially optimizing different variables while keeping others fixed. Herein, we adopt an \gls{iao} based on a \gls{ccm}, which alternately solves the power allocation (\gls{ee} optimization), beamforming, and \gls{ris} phase shift optimization variables for a generic \gls{ris}-aided \gls{m-mimo} system as illustrated in Figure \ref{fig:subfig_d2}. The \gls{iao} optimization steps were implemented as follows:

\noindent {{\bf Step 1}} -- \textit{Fix Power Allocation}: 

\begin{itemize}
    \item[\dm]Initially, the power allocation for \gls{iot} devices is fixed. Using the linear \gls{mmse} receiver ($\boldsymbol{W}_{\text{MMSE}}$), the \gls{bs} combining vectors are determined.
\end{itemize}

\noindent {{\bf Step 2}} --   \textit{Optimize \gls{ris} Phase Shifts ($\boldsymbol{\theta}$)}: 

\begin{itemize}
    \item[\dm]The \gls{ris} phase shifts are then optimized using a manifold-based approach, which respects the complex constraints of the phase shifts.
\end{itemize}

\noindent {{\bf Step 3}} -- \textit{Update \gls{bs} Combining Matrix ($\boldsymbol{W}_{\text{MMSE}}$)}: 

\begin{itemize}
    \item[\dm]After optimizing the phase shifts, the \gls{bs} combining matrix is updated accordingly.
\end{itemize}

\noindent {{\bf Step 4}} -- \textit{Update the Lagrangian multipliers ($\boldsymbol{\lambda}$)}: 

\begin{itemize}
    \item[\dm]Since we apply the Lagrangian relaxation to move the complicated \gls{se} requirement constraint to the \gls{of}, it is necessary to update the Lagrangian multipliers.
\end{itemize}

\noindent {{\bf Step 5}} --\textit{Adjust Device Power ($\boldsymbol{p}$)}: 

\begin{itemize}
    \item[\dm]Finally, the power allocation is optimized, ensuring that each device operates within its power limits while meeting \gls{se} requirements.
\end{itemize}

Comparative results for \gls{ul} transmit power as a function of min-\gls{snr}, number of reflective elements, and noise power at \gls{bs} are illustrated in Figure \ref{fig:subfig_d3}, where \texttt{PSO} represents the PSO-based manifold scheme; \texttt{SD}: steepest-descent for \gls{rm}, and \texttt{CG} holds for the classical conjugate gradient manifolds. Unless stated otherwise, we assume $K = 8$ users are positioned evenly along a semicircle with a radius of 20 meters centered on a \gls{ris}. The distance from the \gls{ris} to the \gls{bs} is 700 meters. Using such a scenario, the distance between users and the \gls{bs} is derived geometrically. The path loss factors between the links are: $\alpha_k$ (UTs-\gls{ris}) = 2, $\beta$ (\gls{bs}-\gls{ris}) = 2.5, and $\gamma_k$ (\gls{bs}-UTs) = 4. Additionally, we use $M = 64$, $N = 100$, $P_{\text{max},k} = 30$ dBm, and $\sigma^2 = -104$ dBm. The \gls{los} channel angles are randomly generated between 0 and $2\pi$. \gls{mcs} are conducted by calculating averages over $10^4$ realizations.

In Figure \ref{fig:subfig_d3}, three scenarios are considered, using the total \gls{ul} transmission power as the metric. In all cases, it is observed that systems without \gls{ris} and systems with \gls{ris} but without optimized reflection phases exhibit the worst conditions, requiring higher $P_{\text{UT}}$ to meet the minimum requirements for all users. Furthermore, in all the considered scenarios, the \texttt{SD} and \texttt{CG} manifold schemes demonstrated nearly identical performance. Notably, without phase shift optimization for the \gls{ris} elements, the performance gain from employing the \gls{ris} is not substantial. Conversely, with both the \texttt{SD} and \texttt{CG} schemes, there is a reduction in $P_{\text{UT}}$ of approximately 50 to 60\%, depending on target values of \gls{se}, number \gls{ris} elements and noise power. Even the PSO-based manifold scheme achieves a power saving of around 40 to 50\%, which is significant given its considerably lower complexity compared to the other optimization schemes.

\subsection{\textbf{Application 3: RIS-Enabled Intra-Cell Pilot Reuse}} 
\label{sec:app3}

Since the inception of \gls{m-mimo} technology, pilot contamination has been the main bottleneck of such systems. The impossibility of allocating orthogonal pilot sequences for every user in the system due to the limited channel coherence blocks leads to the necessity of reusing pilots, which thus results in a directed interference known as pilot contamination. Therefore, a crucial trade-off arises in the system design between the number of orthogonal pilots and the associated overhead for their \gls{ul} transmission, which can severely penalize the system's \gls{se}. As more pilots are reused, the channel estimation overhead improves spectral efficiencies if the associated pilot contamination remains under control.

Given the above scenario, another quite appealing use case for \glspl{ris} is {\it enabling intra-cell pilot reuse} in a \gls{m-mimo} system. Authors in \cite{Marinello24} have shown that, given a set of orthogonal pilots already in use in a cell, each additional \gls{ris} employed at the cell enables the total reuse of such pilot set among the users aided by that \gls{ris}. This is equivalent to having $\tau_p$ orthogonal pilots in a cell with $R$ \glspl{ris}, up to $K = (R+1)\tau_p$ users can be served. This scenario is illustrated in Figure \ref{fig:subfig_m1}. 

\begin{figure}[!htbp]
\begin{subfigure}[b]{1\linewidth}
\centering
\includegraphics[width=1\textwidth]{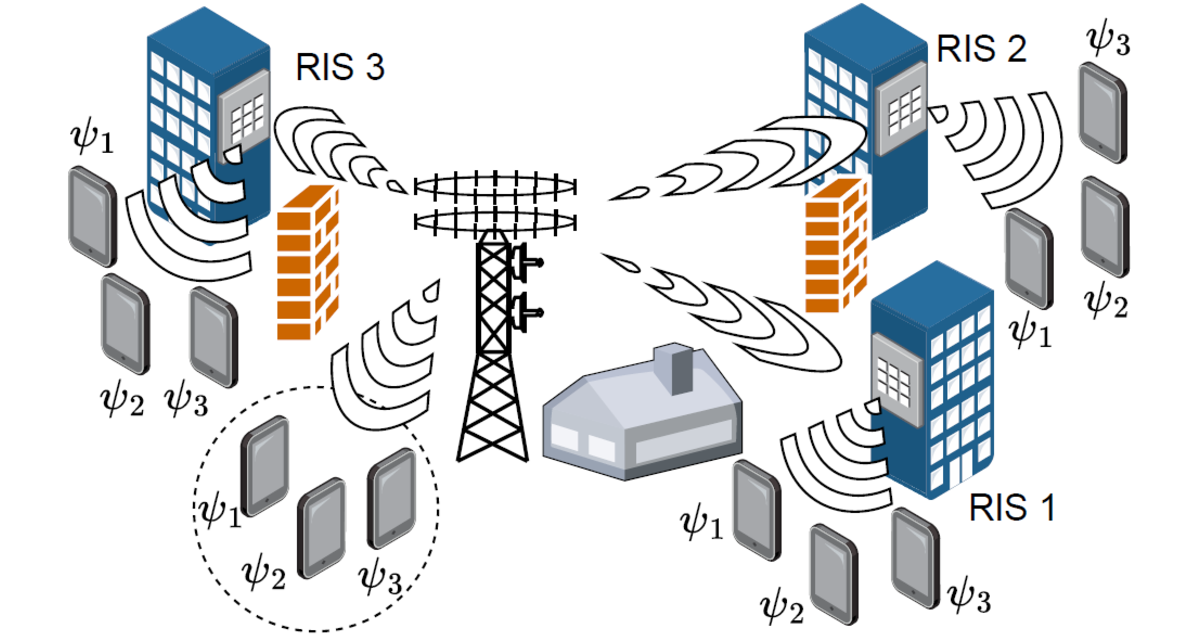}   
    \caption{\gls{mu} \gls{m-mimo} assisted by multiple \glspl{ris}.}
    \label{fig:subfig_m1}
\end{subfigure}

\vspace{2mm}

\begin{subfigure}[b]{1\linewidth}
    \centering
    \includegraphics[width=1\textwidth]{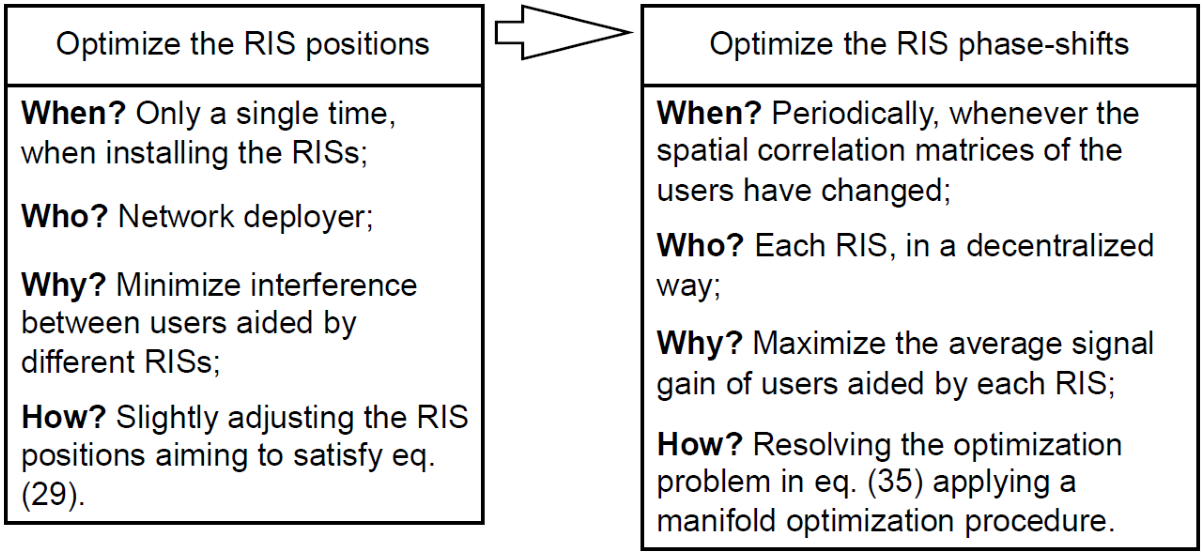}
    \caption{ {\bf Stage 1} \phantom{xxxxxxxxxxxxxxxxxxxxxxxxxxxxxxxxxxxx} \hspace{-1.5cm} {\bf Stage 2}.  }
    \label{fig:subfig_m2}
\end{subfigure}

\vspace{2mm} 

\begin{subfigure}[b]{1\linewidth}
    \centering
    \includegraphics[trim={1mm .1mm 0 0},clip, width=3.4in]{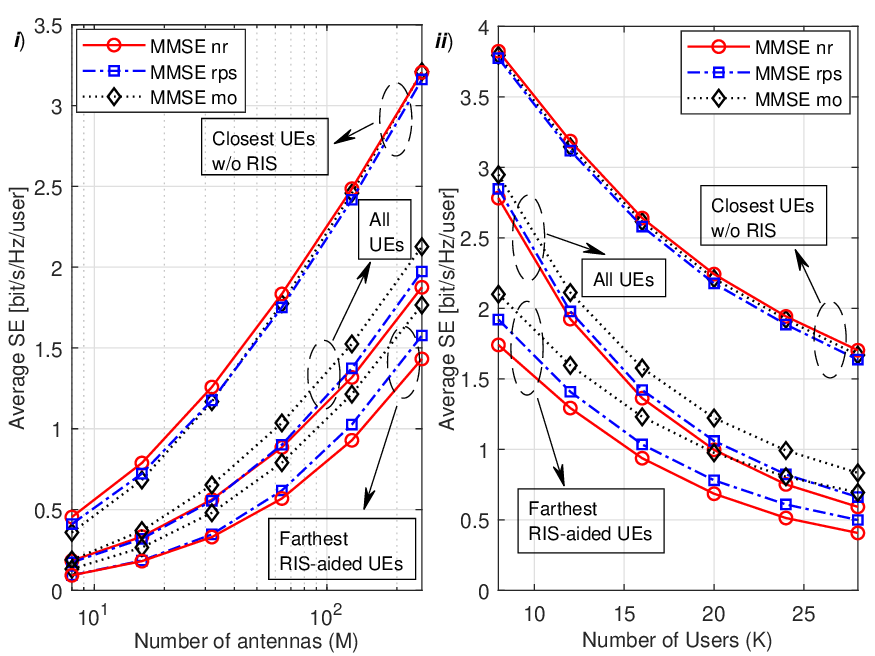}
    \caption{\phantom{xx} \textit{\textbf{i}}. Average SE $\times M$  \hspace{10mm}  \,\, \textit{\textbf{ii}}. Average SE $\times K$}
    \label{fig:subfig_m3}
\end{subfigure}

\caption{ {\textsc{Application 3}:}\small {\bf (a)} \gls{mu} \gls{m-mimo} communication system assisted by multiple \glspl{ris}, each deployed on the facades of buildings. The users in the dotted circle area are served without the aid of any \gls{ris}. They, as well as the users served by each \gls{ris}, share the same set of pilot sequences in our investigated scenario. \, {\bf (b)} Execution tasks of the {\bf two stages \gls{ris}-aided pilot reuse method}; \, {\bf (c)}  \gls{ul} \gls{se} with {$N=256$, $\tau_p=4$, and increasing:  \,\, \textit{\textbf{{i}}}. $M$, when $R=3$, and $K=16$ \glspl{ue};  \,\, \textit{\textbf{{ii}}}. $K$, $R$, and $\varsigma$,} such that $K=\varsigma \, \tau_p = (R+1)\, \tau_p$, when $M=128$ antennas at the \gls{bs}.}
\label{fig:spatial_dist}
\end{figure}

A two-stage methodology is proposed to make it possible while controlling the intra-cell pilot contamination. The first stage consists of optimizing the deployment locations of the \glspl{ris}, with an angular grid being obtained as the optimized positions, leading to reduced interference between the users served by different \glspl{ris}. The second stage consists of optimizing the \gls{ris} phase shifts to maximize the average channel gain of the \gls{ris}-aided users, which is done by applying the \gls{mo} framework based solely on the statistical \gls{csi} of the users. This latter feature has the advantage of requiring less frequent \gls{ris} re-configurations, simplifying channel estimation since the isolated \gls{bs}-\gls{ris} and \gls{ris}-\glspl{ue} channels are not needed, reducing the necessity of deploying active elements at the \gls{ris} side. Figure \ref{fig:subfig_m2} summarizes the main characteristics of each stage.

Figure \ref{fig:subfig_m3}{.\textit{(\textbf{i})} depicts the average \gls{se} performance with the increasing number of \gls{bs} antennas, considering $R=3$, $N=256$, and $K=16$ \glspl{ue} sharing $\tau_p=4$ pilots. It shows the average performances taking into account only the closest \glspl{ue} not aided by any \gls{ris}, only the farthest \gls{ris}-aided \glspl{ue}, and all \glspl{ue} as well. It employs the \gls{mmse} detector while considering as benchmarks the \gls{m-mimo} system without \gls{ris} (denoted as \texttt{nr}), and a scheme employing \gls{ris} with random phase shifts (denoted as \texttt{rps}). On the other hand, the results with the \gls{ris} employing phase shifts obtained via the \gls{mo} approach are denoted by \texttt{mo}.  One can see that the \gls{se} performance of the closest \glspl{ue} {does not suffer significant changes} employing the proposed approach, while the {average \gls{se} per user} performance of the farthest \glspl{ue} is improved {regarding \texttt{nr} and \texttt{rps} approaches}. This occurs since the average channel gain of the \gls{ris}-aided \glspl{ue} increases, {and} the resultant effect is {an} improvement in the \gls{se} when averaged between all \glspl{ue}. The farthest \glspl{ue}' \gls{se} employing $M=128$ antennas increases from 0.929 bit/s/Hz without \gls{ris} to 1.026 bit/s/Hz ($\approx 10\%$) with \texttt{rps} and 1.214 bit/s/Hz ($\approx 31\%$ gain in \gls{se}) with \texttt{mo} {method}. When averaging between all \glspl{ue}, the performance also increases $\approx 16 \%$ with \texttt{mo} in comparison with no \gls{ris} {scheme}.

Then, it is keep fixed the number of pilots as $\tau_p = 4$ and the number of \gls{bs} antennas as $M=128$, while the \gls{prf}  $\varsigma$ increases together with the number of \glspl{ris} and \glspl{ue}, such that $\varsigma = R+1 = K/\tau_p$ holds. Figure \ref{fig:subfig_m3}.(\textit{\textbf{ii}})  
depicts how the \gls{ul} \gls{se} is affected by such aggressive intra-cell pilot-reuse scenarios while showing that the proposed methodology effectively leverages the \glspl{ris} to improve performance in these challenging conditions. One can see an almost linear increase of $\approx 0.3$ bit/s/Hz per \gls{ue} achieved with \texttt{mo} compared with no \gls{ris} {strategy} when averaging between the farthest \gls{ris}-aided \glspl{ue}. As such \glspl{ue} are a fraction of $\frac{\varsigma-1}{\varsigma}$ of the total number of \glspl{ue}; the \gls{ul} \gls{se} gain when averaging between all \glspl{ue} starts from $\approx 0.16$ bit/s/Hz and gradually converges to the same increase of $\approx 0.3$ bit/s/Hz {as $K$ increases}. Besides, if one fixes a target \gls{ul} \gls{se} performance of 1 bit/s/Hz, the number of \glspl{ue} can be increased from 20 to 24 when employing the proposed methodology, with neither an increase in the training overhead nor significant increases in power consumption or processing complexity.

\subsection{\textbf{Application 4: RIS-Aided Grant-Free Random Access for Massive Machine-Type Communication (mMTC)}}\label{sec:app4}

Empowered by the rapid development of \gls{iot} applications, the \gls{mmtc} scenario will play an essential role in the upcoming 6G technology. The enormous number of \gls{mtc} devices usually have sparse activities and low data volumes to transmit, thus requiring new access technologies, which should be decentralized and uncoordinated for scalability. \glspl{ris} can be leveraged to improve the connectivity of the \gls{mmtc} network, improving the {channel} propagation conditions while {simultaneously} managing access to the communication channels, (Figure  \ref{fig:subfig_m21}). \gls{gf} \gls{ra} is a suitable solution for such scenarios, aiming to avoid the excessive overhead of acquiring a grant and performing centralized scheduling procedures. Therefore, another appealing use case for the \glspl{ris} involves developing \gls{ris}-aided \gls{gf} \gls{ra} protocols for \gls{mmtc} systems. 

\vspace{2mm}
\noindent{\bf \gls{gf}-\gls{ra} protocol under single-beam \gls{ris} passive beamforming}: A scheme is proposed in
\cite{Croisfelt23}, which consists of a 2-step protocol. In the first step, the \gls{bs} transmits \gls{dl} pilot signals while the \gls{ris} sweeps its reflection configuration to cover all the devices' areas. The devices receive these reflected pilots and can thus learn which \gls{ris} configuration leads to the highest channel gain for them in the so-called \gls{cs} procedure. In the second stage, the \gls{ris} again sweeps its reflection configurations while the devices transmit their payloads in the time slot corresponding to their chosen \gls{ris} configuration from the previous step. Since the devices are supposed to be uniformly distributed in the covered area, their access in the protocol's second step tends to be uniformly distributed, minimizing collisions and improving channel gains simultaneously.

\begin{figure}[!htbp]
\begin{subfigure}[b]{1\linewidth}
\centering
\includegraphics[width=1\linewidth]{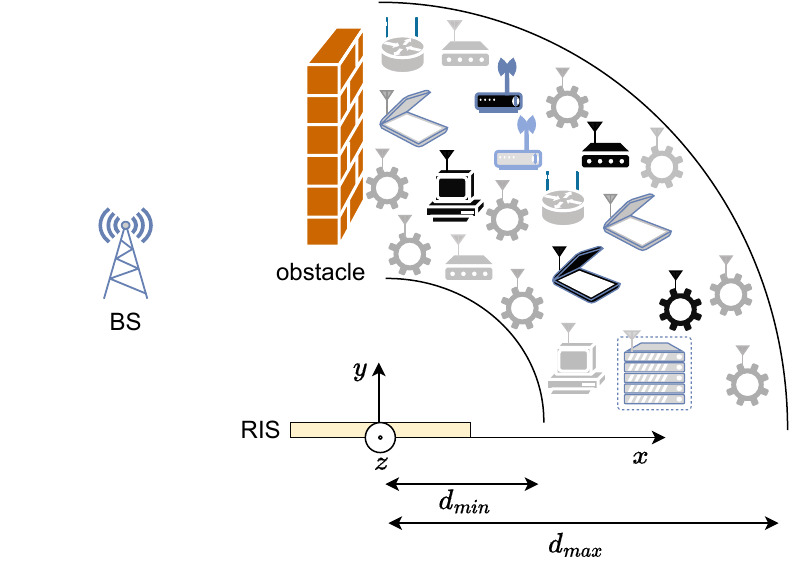}
\caption{\gls{mu} \gls{ris}-aided \gls{m-mimo} scenario.}
\label{fig:subfig_m21}
\end{subfigure}

\begin{subfigure}[b]{1\linewidth}
\centering
\includegraphics[width=1\linewidth]{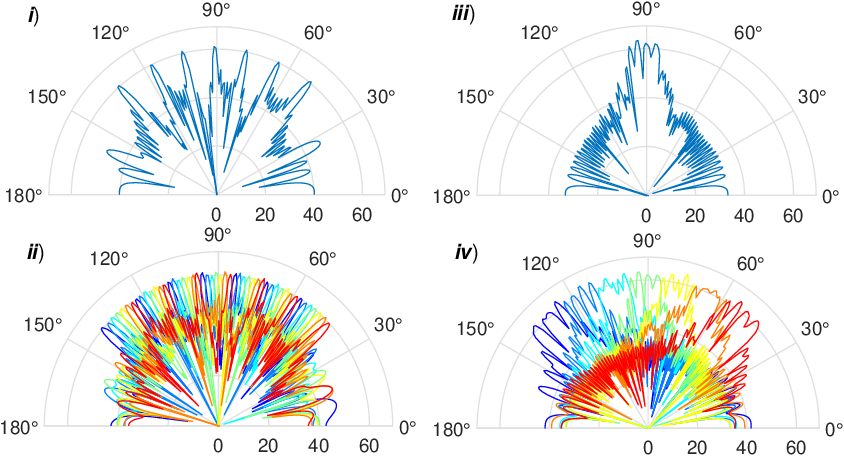}
\caption{\footnotesize{\textit{\textbf{i}}}) Single interleaved multi-beam reflection pattern;
{\textit{\textbf{ii}}}) temporal sequence of interleaved multi-beams covering $[45^o;\, 135^o]$; {\textit{\textbf{iii}}}) a single consecutive multi-beam reflection pattern; {\textit{\textbf{iv}}}) temporal sequence of consecutive multi-beams covering $[45^o;\, 135^o]$}
\label{fig:subfig_m22}
\end{subfigure}

\begin{subfigure}[b]{1\linewidth}
\includegraphics[width=1\linewidth]{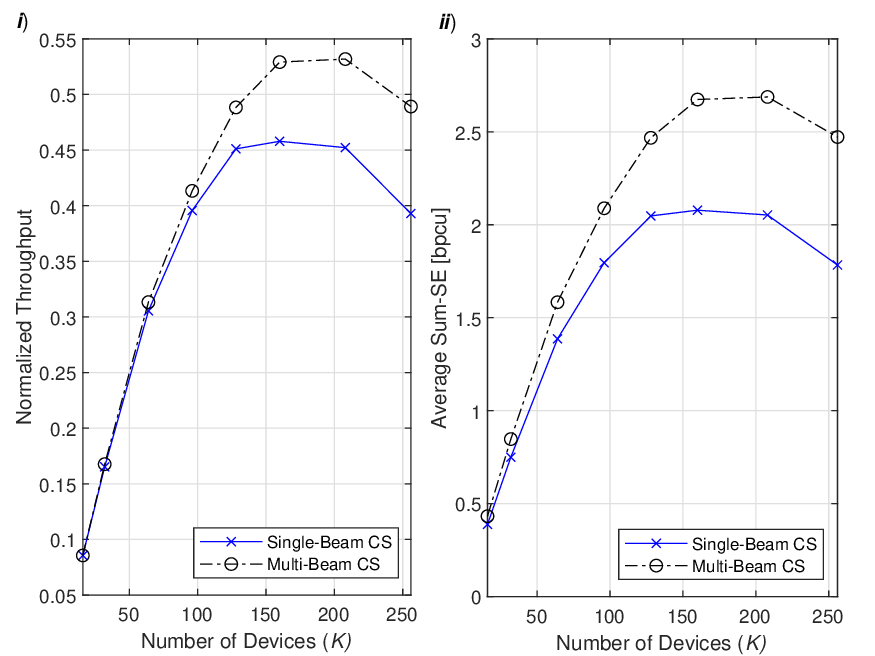}
\begin{minipage}{0.45\linewidth}
\centering
\phantom{xxxxx}{\footnotesize \textit{\textbf{i}}. Normalized throughput.}
\end{minipage}
\hfill
\begin{minipage}{0.45\linewidth}
\centering
{\footnotesize \textit{\textbf{ii}}. Average sum-\gls{se}.}
\end{minipage}
\caption{\gls{ra} performance of the \gls{ris}-aided \gls{gf} protocol with multi-beam channel.}
\label{fig:subfig_m23}
\end{subfigure}

\caption{{\textsc{Application 4}:} \small {\bf (a)} System model;  {\bf (b)} multi-beam reflection patterns designed via \gls{mo}; {\bf (c)} \gls{ra} performance of the \gls{gf} protocol.}
\label{fig:illust}
\end{figure}

\vspace{2mm}
\noindent{\bf \gls{gf}-\gls{ra} protocol under multi-beam \gls{ris} \gls{pbf}}: The single-beam approach can lead to a significant training overhead since many communication resources are usually required in the first step to exhaustively sweep the \gls{ris} configuration among the whole covered area with a single beam per time. To overcome this limitation, a promising solution is to replace the single-beam training approach with a multi-beam one, which should cover the whole devices' area and allow them to find their best \gls{ris} reflection configuration in a reduced training interval. For this sake, the multi-beam \gls{ris} configurations can be designed in two stages: the first one is composed of consecutive beams, which allows the devices to discover to which sector of the covered region they belong; the second one is composed of interleaved beams, and allows the devices to discover in which portion of the sector they are located. However, for the multi-beam reflection design, a simple linear combination of DFT steering vectors does not satisfy the unit modulus constraint and is not easily generated in a passive \gls{ris}. Therefore, an optimization problem can be formulated to obtain a uniform high-gain reflected multi-beam in the intended directions while keeping low reflection gains in the unintended ones, subject to the unit modulus constraint. Once again, the \gls{mo} framework shows to be a promising tool to solve such optimization problems, as discussed in \cite{Wang21}, and can be adapted to design the \gls{ris}-aided \gls{gf}-\gls{ra} protocol. Figure \ref{fig:subfig_m22} depicts the multi-beam reflection patterns obtained by the \gls{mo} framework.

Figure \ref{fig:subfig_m23} evaluates the performance of the \gls{ris}-aided \gls{gf}-\gls{ra} protocols employing single-beam and multi-beam channel sounding approaches. It first compares the normalized throughput performance, defined as the ratio between the number of succeeding devices and available resources. The investigated scenario is composed of a \gls{m-mimo} \gls{bs} with 128 antennas aided by a \gls{ris} with $64 \times 4$ elements in the horizontal and vertical dimensions, respectively. For simplicity, both \gls{bs}, \gls{ris}, and devices are assumed to be in the same horizontal plane; therefore, up to 64 DFT-based orthogonal beams could be used, among which only 46 falls in the considered devices region between $45^o$ and $135^o$, which are those employed in the access stage (step 2). In each access stage, the devices also transmit a random pilot sequence of length $\tau_p = 4$, which may allow them to successfully access the same beam provided they choose different pilots and achieve an \gls{sinr} above the decoding threshold. Thus, the normalized throughput in this scenario is the number of succeeding devices divided by 184, which is the total number of resources combining beams and pilots in the access stage. As can be seen from Figure 
\ref{fig:subfig_m23}.({\it \textbf{i}}), the multi-beam \gls{cs} approach achieves a maximum normalized throughput of $0.5318$, which, compared to $0.4579$ achieved by the single-beam one, represents a remarkable improvement of 16.14\%. On the other hand, Figure 
\ref{fig:subfig_m23}.({\it \textbf{ii}}) evaluates the sum-\gls{se} of the network, which also accounts for the lower overhead required by the multi-beam \gls{cs} approach. While the single-beam \gls{cs} approach needs 46-time slots in step 1 in the evaluated scenario, the multi-beam approach can evaluate step 1 with only 14-time slots. This is accomplished by expanding the targeted area in 49 beams, scanned in 7 configurations of 7 consecutive multi-beams, followed by another 7 configurations of 7 interleaved multi-beams, as depicted in Figure \ref{fig:subfig_m22}. In this way, the multi-beam \gls{cs} approach achieves a maximum sum-\gls{se} of $2.6873$ bpcu, which, in comparison with $2.0783$ bpcu achieved by the single-beam one, represents an important improvement of 29.3\%. Besides, it is worth mentioning that there is room for larger gains if the reflection patterns are further improved (reducing side-lobe levels and beam ripple), and/or if the multi-beam reflection approach is also leveraged in the access stage.


\section{CONCLUSION} \label{sec:VII}

\gls{mo} provides a structured and powerful approach to tackling complex optimization problems in wireless communication systems, especially in \gls{ris}-aided \gls{m-mimo} systems scenarios for next-generation wireless communications, offering significant advantages over conventional optimization methods. By leveraging the geometric properties of the manifold, we can achieve more efficient and effective solutions compared to conventional optimization methods.  \gls{mo} methods offer significant advantages over traditional optimization methods in wireless communications, particularly in handling non-convex constraints and high-dimensional data. While heuristic evolutionary algorithms, convex relaxation techniques, and gradient-based methods each have their merits, \gls{mo} methods provide a more natural and efficient framework for solving complex optimization problems in this domain. The key advantages of \gls{mo} methods include their ability to exploit geometric properties, manage high-dimensional data, and handle various constraints and symmetries, making them particularly well-suited for next-generation wireless communication systems.

We have explored in details five applications of the \gls{mo} framework aiming to optimizing \gls{ris}-aided \gls{m-mimo} wireless communication systems; we have proposed: {a})  an efficient \gls{mo}-based algorithm for optimizing the beamforming weights and \gls{ris} phase shifts, maximizing the sum of received \glspl{sinr}; {b}) maximize the minimum rate of the network by leveraging \gls{rcg} within a \gls{ccm}; 
{c}) an \gls{rm}-based \gls{ris} phase shift optimization method aimed at minimizing the \gls{ul} transmit power in \gls{iot} networks with \gls{ris};
{d}) \gls{ris} phase shifts optimization method for intra-cell pilot reuse with statistical \gls{csi} using \gls{mo} methodology, highlighting remarkable performance improvements in both \gls{ul} and \gls{dl} transmissions; 
{e}) multi-beam design optimization problem in a \gls{ris}-based \gls{gf} random access protocol. 
A key advantage of the \gls{mo} approach is the ability to handle non-convex constraints naturally. These applications illustrate the substantial benefits of using \gls{mo} techniques to enhance the performance, fairness, and efficiency of \gls{ris}-aided \gls{m-mimo} systems.

\bibliographystyle{bibliography/IEEEtranIES}
\bibliography{bibliography/bib_Manifolds}

\end{document}